\documentclass{svjour3}                     
\smartqed  
\usepackage{graphicx}

\usepackage{placeins}

\usepackage{hyperref}

\usepackage{color}
\usepackage{amsmath}
\usepackage{amssymb}

\definecolor{azul}{RGB}{33,72,195}
\definecolor{granate}{RGB}{195,33,72}
\definecolor{morado}{RGB}{153, 102, 255}

\begin{document}

\title{Explainable artificial intelligence toward usable and trustworthy computer-aided diagnosis of multiple sclerosis from Optical Coherence Tomography
\thanks{This work was partially supported by the national research grants PID2019-104358RB-I00 (DL-Ageing project),
Government of Aragon Group Reference $T64\_20R$ (COS2MOS research group),
Carlos III Health Institute grants PI17/01726 and PI20/00437, and by the Inflammatory Disease Network
(RICORS) (RD21/0002/0050) (Carlos III Health Institute).
Monica Hernandez is with the Computer Sciences Department and the Aragon Institute on Engineering Research,
University of Zaragoza, Spain.
(e-mail: mhg@unizar.es).
}
}


\titlerunning{xAI toward usable and trustworthy CA diagnosis of MS from OCT}        

\author{Monica Hernandez, Ubaldo Ramon-Julvez, Elisa Vilades, Beatriz Cordon, Elvira Mayordomo, Elena Garcia-Martin}

\authorrunning{Hernandez et al.} 

\institute{M. Hernandez \at
              University of Zaragoza \\
              Tel.: +34-876-865544\\
              Fax: +34-976-761914\\
              \email{mhg@unizar.es}           
}

\date{Received: date / Accepted: date}

\maketitle

\begin{abstract}
{\it Background:} Several studies indicate that the anterior visual pathway provides information about
the dynamics of axonal degeneration in Multiple Sclerosis (MS).
Current research in the field is focused on the quest for the most discriminative features among patients and controls
and the development of machine learning models that yield computer-aided solutions widely usable in clinical practice.
However, most studies are conducted with small samples and the models are used as black boxes.
Clinicians should not trust machine learning decisions unless they come with comprehensive
and easily understandable explanations.
{\it Materials and methods: } A total of 216 eyes from 111 healthy controls and 100 eyes from 59 patients with relapsing-remitting
MS were enrolled.
The feature set was obtained from the thickness of the ganglion cell layer (GCL) and the retinal nerve fiber layer (RNFL).
Measurements were acquired by the novel Posterior Pole protocol from Spectralis Optical Coherence Tomography (OCT) device.
We compared two black-box methods (gradient boosting and random forests) with a glass-box method (explainable boosting machine).
Explainability was studied using SHAP for the black-box methods and the scores of the glass-box method.
{\it Results: }
The best-performing models were obtained for the GCL layer.
Explainability pointed out to the temporal location of the GCL layer that is usually broken or thinning in MS and the relationship
between low thickness values and high probability of MS, which is coherent with clinical knowledge.
{\it Conclusions: } The insights on how to use explainability shown in this work represent a first important step toward
a trustworthy computer-aided solution for the diagnosis of MS with OCT.


\end{abstract}


\section{Introduction}


\subsection{Multiple Sclerosis}

Multiple sclerosis (MS) is a neurodegenerative disease affecting the central nervous system (CNS).
In MS the immune system attacks nerve fibers and myelin sheathing in the brain and spinal cord.
The consequences are inflammation, demyelination, and axonal degeneration in the whole CNS,
destroying nerve cell processes and altering the electrical messages in the brain.
A confirmed diagnosis of MS is difficult, especially in the early stages of the disease when symptoms could be minor,
sporadic, or even resemble other disease conditions.
The diagnosis is based on the McDonald criteria, consisting of clinical, radiographic, and laboratory
parameters extracted from a neurological examination and the history of neurological symptoms~\cite{Thompson_18}.
The initial version of the McDonald criteria was proposed in 2001 and it has been revised multiple times.
Most recent criteria date from 2017.
Tests typically used for the neurological evaluation include:

\begin{itemize}
\item Magnetic Resonance Imaging of the brain and/or the spinal cord looking for plaques or scars typical of MS.
\item Cerebrospinal fluid evaluation using a lumbar puncture looking for immunological abnormalities.
\item Evoked potential studies measuring the electrical impulses along pathways in the brain and the spinal
cord or along the optic nerve in the case of patients showing optic neuritis.
\end{itemize}
\noindent These tests are time-consuming, costly, and invasive.
For example, magnetic resonance devices are expensive, and image acquisition may range from 10 to 30 minutes.
Cerebrospinal fluid samples are acquired through a lumbar puncture that can be described as unpleasant and painful.
Evoked potential tests requires the acquisition of data through about 2 hours.

\subsection{MS and ophthalmic region deterioration}

In the last decades, several studies indicate that the anterior visual pathway provides information about the
dynamics of axonal degeneration in MS~\cite{Petzold_10,GarciaMartin_10,Manogaran_19,Britze_18}.
The hypothesis is that MS may be early diagnosed and then followed up from the quantification of the axonal damage in the
neuroretina~\cite{GarciaMartin_10,Balk_16}.
The optic nerve is being extensively investigated as a potential location in future updates to the McDonald criteria~\cite{Toosy_22}.

Optical coherence tomography (OCT) is a fast (order of seconds) and non-invasive imaging technique that allows
the segmentation and quantification of the thickness of the different neuroretinal layers with less than 2 microns
of precision.
The discomfort for the patient during the acquisition is minimum, since he/she only has to stare at a fixed point for a few
seconds while the images are acquired.
In particular, OCT has revealed that the retinal nerve fiber layer (RNFL) reflects the axonal integrity,
while the ganglion cell layer (GCL) shows a progressive thinning since the early stages of MS~\cite{Petzold_10,Britze_17,Wicki_18}.
These findings point out OCT-derived measurements as biomarkers potentially useful in the diagnosis of the disease.

Commercial OCT devices are distributed with accurate retinal layer segmentation algorithms.
They also provide quantification of the layer thicknesses in different ways and locations.
One-dimensional measurements of the layer thicknesses around the optic nerve head are taken using
circular sections such as the peripapillary circle scan. 
Using this protocol, the thickness of the peripapillary RNFL (pRNFL) in the temporal sector has
been associated with a level of cognitive and physical disability in MS~\cite{Petzold_10,Birkeldh_17}.
In its wide scan protocol, Triton OCT device (Topcon, Tokyo, Japan) provides a two dimensional
grid of measurements of the thickness in macular and peripapillary areas (45 $\times$ 60 points).
Posterior Pole (PPole) is a new two-dimensional quantification protocol for Spectralis OCT devices
(Heidelberg Engineering, Germany).
Similarly to Triton super grid, PPole protocol quantifies the different layer thicknesses in an $8 \times 8$ grid.
PPole has been satisfactorily used in early glaucoma assessment~\cite{Zang_17,Casado_19} and recent studies are conducted
to establish its potential to detect changes in RNFL and GCL due to MS~\cite{Martucci_21,Vilades_23}.
Current research in the field is focused on the quest for the most discriminative measurements among MS and
controls.

In particular, the PPole protocol of Spectralis OCT constitutes a significant advance in measurement acquisition,
because it is based on the high reliability provided by applying the Anatomical Positioning System (APS) to OCT imaging.
The APS works by locating points on the
eye using two fixed structural landmarks: the centre of the fovea and the centre of the Bruch membrane opening. This
allows much more precise monitoring, detecting even the slightest alteration or inter-eye difference because using the APS,
in conjunction with the True Track eye-tracking system (TruTrack, Heidelberg Engineering), ensures accurate identification
of macula position in each eye based on head tilt and eye cyclotorsion. With other protocols, part of the differences observed
in the measurements could be due to differing alignment between the right and left eyes, to minor anatomical or refractive differences
between them, or even to the patient's head being positioned on the OCT device's chin rest at a slight angle imperceptible to the clinician.

\subsection{Machine learning in the diagnosis of MS. Usable solutions.}

The different ways of quantifying the layer thicknesses have been used as feature sets
for building machine learning systems toward the computer-aided diagnosis of MS.
The comprehensive review recently provided by Aslam et al. in~\cite{Aslam_22} describes
the advances in the performance of these methods and discusses some limitations.

Garcia-Martin et al.~\cite{GarciaMartin_13} pioneered the studies of the discrimination capability of multiple
sclerosis vs healthy control (MS vs HC) subjects using ML systems.
In this work, artificial neural networks (ANN) were used with measurements from the circumpapillary RNFL
thickness from Spectralis SD-OCT.
The study involved 106 MS patients and 115 age-matched HC.
The best-performing model obtained an AUC of 0.94.

Palomar et al.~\cite{Palomar_19} studied the diagnostic capacity of traditional machine learning methods
in different locations of the neuroretina, centered on the macula (macular protocol) or on the optic nerve (peripapillary protocol).
The authors used a two-dimensional grid of measurements from Triton SS-OCT device.
The thickness was computed for the RNFL and GCL+ (GCL + IPL layers, IPL standing for the inner plexiform layer).
The study involved 80 MS patients and 180 age-matched HC.
RNFL showed as the best feature set for the prediction of the disease with a 97.24\% of accuracy with decision trees (DT).

Cavaliere et al.~\cite{Cavaliere_19} combined support vector machines (SVM) with the mean values over right
and left eyes of the macular thickness and the peripapillary area layer thickness from Triton SS-OCT.
Feature selection was performed and the models were built over a feature set of dimension three.
The most discriminant values were the total GCL++ thickness (between the inner limiting membrane to the inner
nuclear layer boundaries)
and the macular retina thickness in the nasal quadrant of the outer and inner rings.
The study involved 48 MS and 48 HC subjects.
The best-performing model obtained a 91\% of accuracy.

Montolio et al.~\cite{Montolio_21} performed an exhaustive study involving different ML methods in the diagnosis
of MS from the RNFL thickness averaged over the peripapillary, superior, nasal, inferior, temporal, and foveal
areas.
The images were acquired with a Cirrus high definition OCT device.
The study involved 108 MS and 104 HC subjects.
The study also included a temporal analysis of the evolution of the measurements with a follow-up of 10 years.
The best-performing model was obtained from a model ensemble of the best-performing models generated in the
cross-validation.
The accuracy of the ensemble was 87.7 \% while kNN reached an accuracy of 85.4 \% and SVM of 84.4 \%.

Garcia-Martin et al.~\cite{GarciaMartin_21}, proposed a method for the early diagnosis of MS.
The authors used a two dimensional grid of measurements from Triton OCT device acquired using wide protocol (12 $\times$ 9 mm).
The thickness was computed for the RNFL, GCL+, and GCL++ layers and features were extracted using a thresholding
over the Cohen's d metric.
The study involved 48 MS patients and 48 healthy controls.
GCL++ layer showed to be the one with the highest discriminant capacity, with a 98.00\% of accuracy in this case
with a shallow neural network (NN).
It should be noticed the contradiction with Palomar et al. findings~\cite{Palomar_19}.

Montolio et al.~\cite{Montolio_22} reproduced the study in~\cite{Montolio_21} with the measurements of the
peripapillary circle scan in the RNFL layer and a different population of subjects.
The study involved 72 MS patients and 30 HC.
The authors reported kNN as the best-performing method with a 95\% of accuracy on a validation set balanced
with SMOTE oversampling method.

Lopez Dorado et al.~\cite{LopezDorado_22} implemented a system based on a convolutional neural network (CNN)
using the two-dimensional grid of measurements from Triton OCT device acquired using wide protocol.
The study involved 48 MS patients and 48 healthy controls.
The dataset was augmented using a generative adversarial network (GAN) and Cohen's d metric was used to select
the retinal structures with the greatest discriminant capacity (resulting in GCL++, complete retina, and GCL+).
The final model obtained a 100\% of accuracy in a leave-one-out evaluation.

The major criticism from Aslam et al. in~\cite{Aslam_22} is that these studies have been conducted
with a very small number of samples. Consequently, the models may not
be robust presenting a high variance. In addition, the models may be biased.
We agree with these appreciations.
Indeed, the population sample and
utilized features are usually
quite different among the studies. This makes it unfeasible to establish which could be the best combination
of feature set and learning system for the task.
Some of these works show comparative tables of the performance of the different methods when they cannot
be objectively compared.
These problems are not exclusive to this area of research and they are found more frequently than is
desirable in different applications of ML in clinical contexts~\cite{Marinescu_20,Hernandez_22}.


The valuable information from these works relies on which features can be potential biomarkers for which
stages of the disease, but the proposed systems are still far from computer-aided solutions usable for the
general clinical practice.
To ensure the reliability of the proposed systems, the models should be built on a large dataset,
demonstrate a high generalization capability, and make decisions coherent with clinical knowledge.
However, this is not a simple task in the great majority of clinical applications and this one is no exception.

The only study to date involving a large dataset has been recently conducted by Kenney et al.~\cite{Kenney_22}.
The authors used Cirrus high-definition OCT and converted Spectralis HP-OCT measurements of the pRNFL and
GCL+ layers.
The absolute difference between eyes and the average between eyes were compared as features.
The study involved 1568 MS patients and 552 HC.
The authors reported the best results for the GCL+ layer with a logistic regression, obtaining an AUC of 0.77.
The AUC obtained for the pRNFL measurements was 0.65.
These results were improved up to an AUC of 0.89 with a composite score combining OCT measurements with
the score of an ophthalmologic test (a not anatomical feature).
These results are far from the impressive performances over the 90\% provided in the small-dataset studies,
but it is not immediate to assess on the causes of the divergence of these results.
%



\subsection{Explainable AI. Trustworthy solutions.}

It is well-known that clinical practitioners do not rely on ML approaches despite their high
accuracy~\cite{ElSappagh_21}.
Most ML methods are black boxes. This means that it is not comprehensive or easily understandable
why or how the system reached a specific decision or whether the system is making decisions with arguments that are
coherent with medical knowledge.
In consequence, medical experts usually do not trust ML decisions unless they come with comprehensive and
easily understandable explanations.

Attending the taxonomy of explainable methods, the trade-off between performance and explainability is known.
There is an inverse relationship between the accuracy of a machine-learning algorithm and our ability to
understand and interpret its output.
Therefore, one may engage the clinical experts with glass-box methods at the expense of decreasing the
accuracy.

In the last years, the novel discipline of explainable AI (xAI) has been developed with interesting methods that try to
open the black box of the most complex methods and provide explanations of their decisions.
Among them, SHAP~\cite{Merrick_20} (\url{https://shap.readthedocs.io}) is an xAI method based on game theory that provides information
among the relationship between the feature values and the predictions of an ML system.
SHAP was proposed with many different ways to graphically represent the SHAP values.
Some of them allow the global identification of the most important features involved in the predictions
and assess whether high or low feature values contribute to high or low probability predictions for a given class.
For particular cases, the SHAP values allow knowing which features contributed to their correct or wrong
classification.

Another approach in xAI consists of the proposal of inherently explainable methods that are performance competitive
with traditional black boxes, breaking the performance-explainability trade-off.
Explainable Boosting Machine (EBM) (\url{https://interpret.ml}) has been recently proposed as a glass-box method
intended to achieve an accuracy similar to competing black-box methods.
Explanations from EBM can be graphically represented in a similar way to SHAP.

The information provided by the different xAI approaches can be analyzed with clinical experts to find out whether
the ML models are making decisions coherent with clinical knowledge, detect biases, or even point out flawed
experimental designs.
Thanks to SHAP and EBM, xAI is nowadays in a mature state to bridge the gap between models generated in an academic
research environment and their effective utilization in clinical practice as usable and trustworthy solutions.

\subsection{Our contribution to the state of the art}

The objective of our work is to study the potential of the PPole protocol from Spectralis SD-OCT combined with ML methods
for the diagnosis of MS with a novel approach based on xAI.
As feature sets, we consider two different ways of representing the measurements of the GCL and RNFL thickness.
The ML methods have been selected among those that have obtained the best performances in other medical applications~\cite{Marinescu_20,Hernandez_22}.
Explainability has been provided using SHAP and EBM.
The models have been built using a local database with the problems usually found in clinical practice:
small size, unbalancing, heterogeneity, etc.
For a better insight on the capabilities of our systems,
we are showing the average and variance of the performance measurements for the different ML methods, the different combinations
of measurement representation, and sample selection.
We are describing how to provide detailed global and local explanations for the ML decisions and how can they be interpreted in
an understandable way.
We analyze whether the obtained explanations are coherent with clinical knowledge.
Finally, the local explanations of the failure cases provide interesting insights into the decision-making process of the
different methods.
We believe that the explainability study shown in this work represent a first important step toward a trustworthy computer-aided
solution for the diagnosis of MS that may help neurologist to get an earlier definitive diagnosis in patients with clinical suspect
of MS or even to get an estimation of the risk of presenting the disease in subjects with unspecific neurological symptoms or with
familiar antecedents of MS.

\subsection{Manuscript organization}

In the following, Section~\ref{sec:Data} describes the data used in our study.
Sections~\ref{sec:Methods} and~\ref{sec:xAI} present the ML and xAI methods used in our work for the discrimination of MS and HC subjects.
Section~\ref{sec:Implementation} describes the experimental design of our study.
Section~\ref{sec:Results} evaluates the models obtained with our data and presents and discusses the explainability results.
Finally, Section~\ref{sec:Discussion} gathers the most remarkable conclusions of our work.


\section{Dataset description}
\label{sec:Data}


The clinical procedures in this study adhered to the tenets of the Declaration of Helsinki.
The experimental protocol was approved by the Ethics Committee of the Miguel Servet Hospital (CEICA), 
Zaragoza, Spain,
and all participants provided written informed consent to participate in the study.
We included patients with definite relapsing-remitting MS (RR MS). Diagnosis was based on the 2010 revision of the
McDonald Criteria and confirmed by a specialized neurologist.


Our dataset comprises a total of 316 eyes from 111 healthy controls and 59 patients of remitting-relapsing MS (RR MS).
The dataset used in~\cite{Vilades_23} is contained into ours.
Table~\ref{table:Demographics} shows the demographics of our data.
Eyes with previous episodes of optic neuritis were excluded in order to avoid bias when assessing how neurodegeneration is 
appreciable in the retinal layers.
Eyes longer than 25.2 mm and refractive errors $\geq$ 5 diopters (D) of spherical equivalent or $\geq$ 3 D
of astigmatism were excluded from the study. Also, an ophthalmologic examination was used to detect any ocular
alterations such as macular or disc optic damage (e.g. subclinical glaucoma) or cataract of media opacity that could affect
functional vision or captured images.

\begin{table}[!t]
\begin{center}
\scriptsize
\vspace{0.25 cm}
\begin{tabular}{|c|c|c|c|}
\hline
& Healthy Controls & Multiple Sclerosis & p-value \\
\hline
\hline
Subjects & 111 & 59 & n.a. \\
Left eyes & 109 & 50 & n.a.\\
Right eyes & 107 & 50 & n.a.\\
Age (years) & 44.74 $\pm$ 11.09 & 41.94 $\pm$ 13.85 & 0.05 \\
Male/Female & 25/86 & 10/49 & n.a. \\
\hline\hline
Z1 GCL thickness & 41.78  $\pm$  2.89 & 36.70  $\pm$  6.62 & $\mathbf{ < 0.001}$\\
Z2 GCL thickness & 49.22  $\pm$  3.68 & 43.50  $\pm$  7.71 & $\mathbf{ < 0.001}$ \\
Z3 GCL thickness & 28.52  $\pm$  2.36 & 27.56  $\pm$  3.40 & $\mathbf{ 0.015}$ \\
Z4 GCL thickness & 27.23  $\pm$  2.46 & 25.42  $\pm$  2.55 & $\mathbf{ < 0.001}$ \\
Z5 GCL thickness & 28.46  $\pm$  2.35 & 26.38  $\pm$  2.87 & $\mathbf{ < 0.001}$ \\
Z6 GCL thickness & 28.39  $\pm$  2.62 & 27.13  $\pm$  3.43 & $\mathbf{ 0.02}$ \\
\hline
Z1 RNFL thickness & 31.87  $\pm$  4.26 & 30.12  $\pm$  5.69 & $\mathbf{ 0.031}$ \\
Z2 RNFL thickness & 28.32  $\pm$  2.99 & 26.66  $\pm$  4.17 & $\mathbf{ 0.004}$ \\
Z3 RNFL thickness & 63.94  $\pm$  10.25 & 60.58  $\pm$  12.09 & $0.070$ \\
Z4 RNFL thickness & 80.37  $\pm$  12.93 & 72.40  $\pm$  13.40 & $\mathbf{ < 0.001}$ \\
Z5 RNFL thickness & 30.46  $\pm$  4.38 & 28.63  $\pm$  4.88 & $\mathbf{ 0.018}$ \\
Z6 RNFL thickness & 22.39  $\pm$  2.73 & 22.91  $\pm$  5.66 & 0.431 \\
\hline
\end{tabular}
\end{center}
\vspace{0.1 cm}
\caption{\small Demographics of the data used in our work.
In addition, we show whether statistical significance can be obtained for the zone-based
Posterior Pole (PPole) measurements of ganblion cell layer (GCL) and retinal nerve fiber layer (RNFL) between our
populations of healthy controls (HC) and multiple sclerosis (MS) subjects.
Bold numbers indicate p values $<0.05$.
}
\label{table:Demographics}
\end{table}

\begin{figure}[!t]
\centering
\scriptsize
\begin{tabular}{c}
\includegraphics[width=0.95\textwidth]{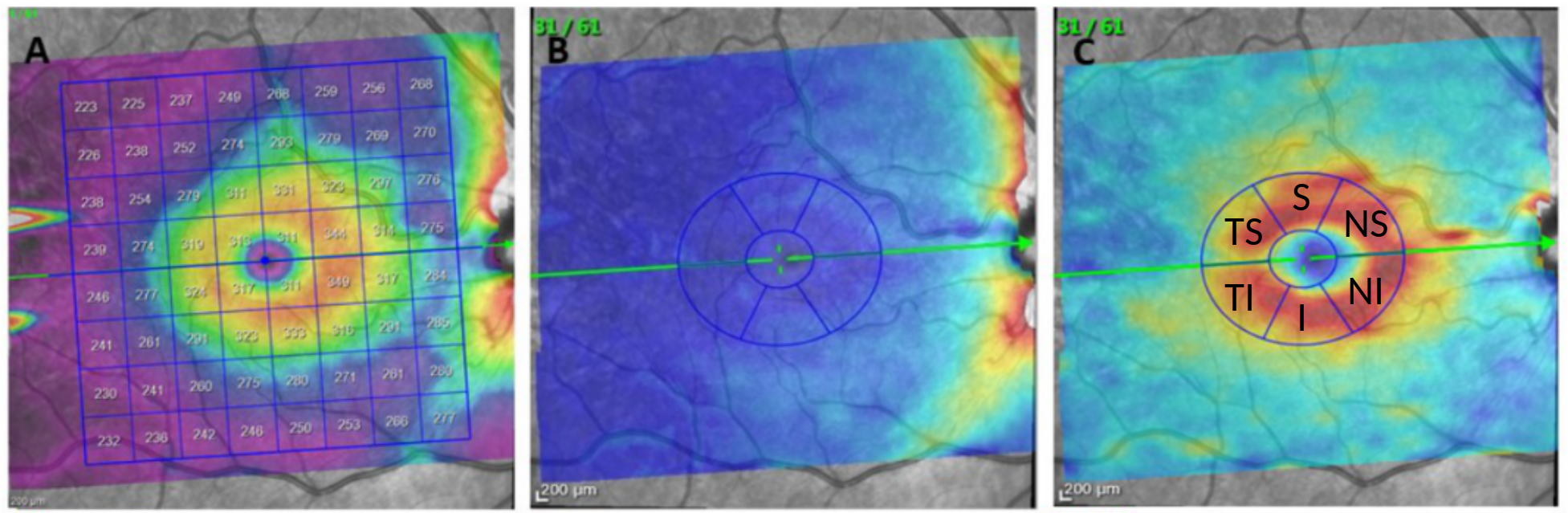} \\
\end{tabular}
\caption{Illustration of the Posterior Pole (PPole) protocol.
Left, standard positioning of the $8 \times 8$ grid centered over the macula.
Center, color thickness map of the ratinal nerve fiber layer (RNFL).
Right, color thickness map of the ganglion cell layer (GCL).
The green line indicates the direction of Spectralis Anatomical Positioning System (APS).
The blue sectors indicate the distribution of superior (S), nasal superior (NS), nasal inferior (NI),
inferior (I), temporal inferior (TI), and temporal superior (TS) areas.
}
\label{fig:PPole}
\end{figure}

Structural measurements of the retina were obtained using the Spectralis OCT device (Heidelberg Engineering, Germany).
The Posterior Pole protocol (PPole), recently included in the device software, was used for all subjects (see Figure~\ref{fig:PPole}).
This protocol incorporates the Anatomic Positioning System (APS), which describes a horizontal line between the fovea and
the entrance of Bruch’s Membrane opening.
Based on this reference line 61 parallel explorations are performed inside an area of $25^\textnormal{o} \times 30^\textnormal{o}$.
APS combined with the True Track system for eye tracking ensures an accurate position of the macula for each individual based
on their head tilt and eye cyclotorsion.

The OCT images were obtained by a single operator blinded to group classification.

\begin{figure}[!t]
\centering
\scriptsize
\begin{tabular}{c}
\includegraphics[width=0.45\textwidth]{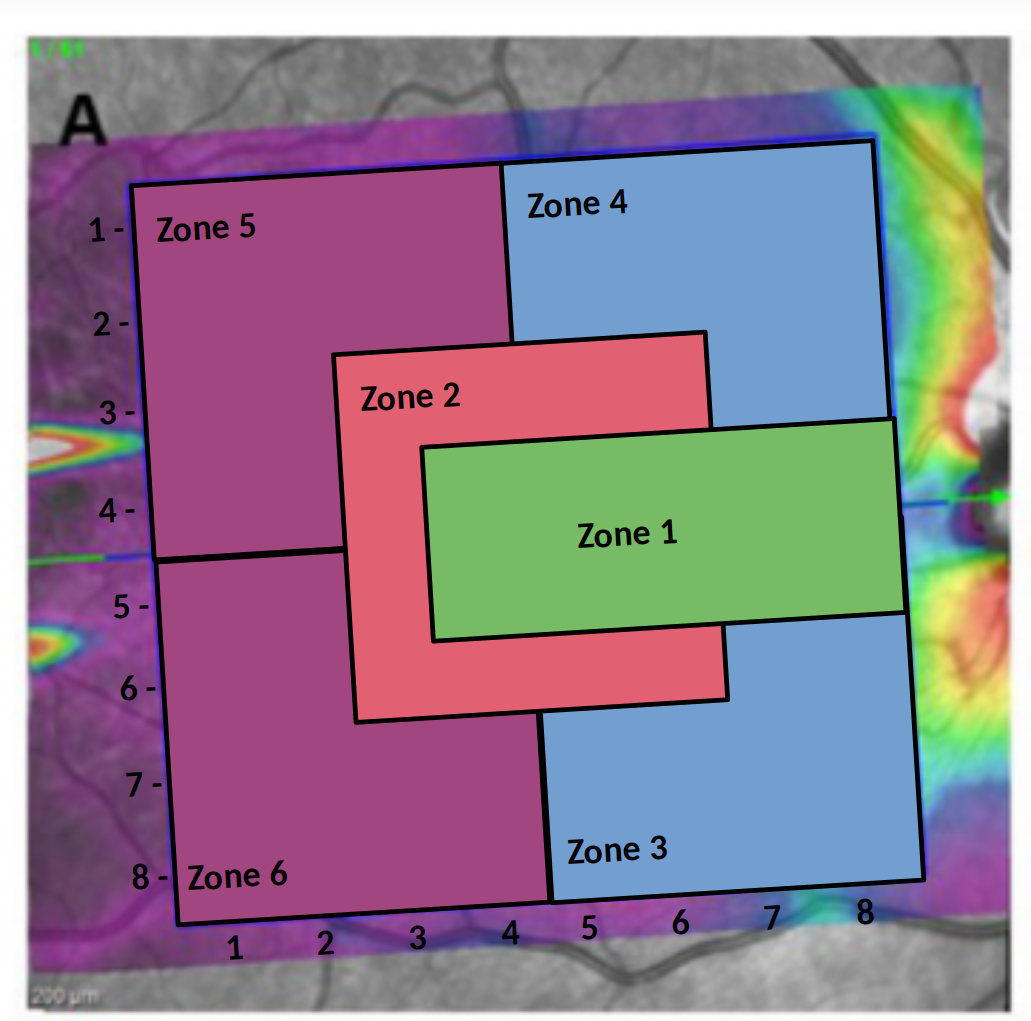} \\
\end{tabular}
\caption{Coordinate system and zone grouping of the 8 $\times$ 8 Posterior Pole (PPole) grid.
In our work we use a coordinate system with origin located at point (1,1) of the PPole grid.
This way, the coordinate naming convention follows the one used by Spectralis optical coherence tomography device.}
\label{fig:PPoleZones}
\end{figure}

\begin{figure}[!t]
\centering
\scriptsize
\vspace{0.25 cm}
MS samples, GCL layer \\
\vspace{0.25 cm}
\begin{tabular}{cccccc}
& subject 1 & subject 2 & subject 3 & subject 4 & subject 5 \\
L & \includegraphics[width=0.13\textwidth]{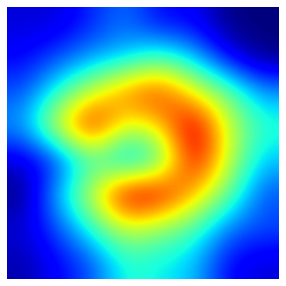} &
\includegraphics[width=0.13\textwidth]{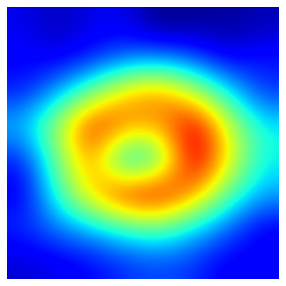} &
\includegraphics[width=0.13\textwidth]{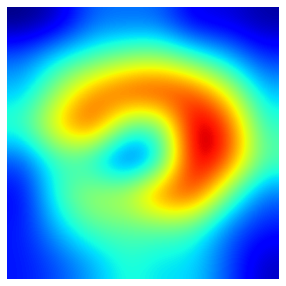} &
\includegraphics[width=0.13\textwidth]{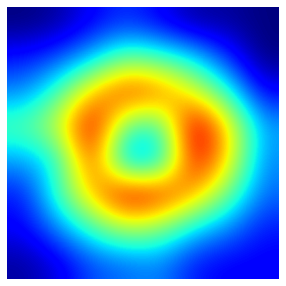} &
\includegraphics[width=0.13\textwidth]{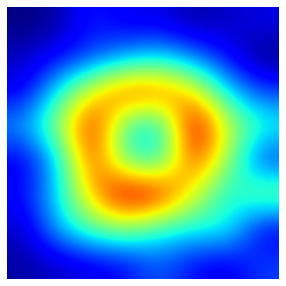} \\
R & \includegraphics[width=0.13\textwidth]{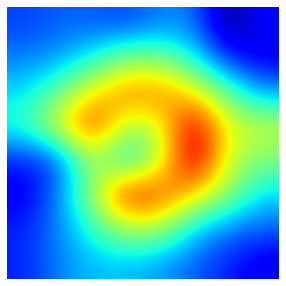} &
\includegraphics[width=0.13\textwidth]{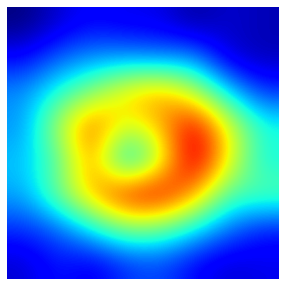} &
\includegraphics[width=0.13\textwidth]{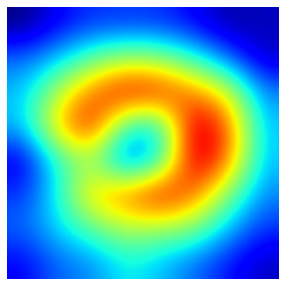} &
\includegraphics[width=0.13\textwidth]{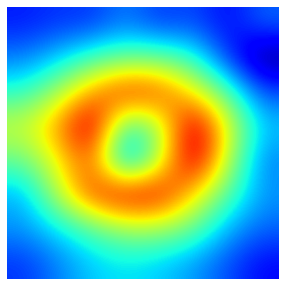} &
\includegraphics[width=0.13\textwidth]{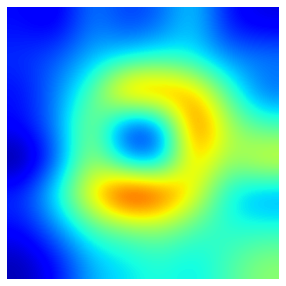} \\
\end{tabular} \\
\vspace{0.25 cm}
HC samples, GCL layer \\
\vspace{0.25 cm}
\begin{tabular}{cccccc}
& subject 6 & subject 7 & subject 8 & subject 9 & subject 10 \\
L & \includegraphics[width=0.13\textwidth]{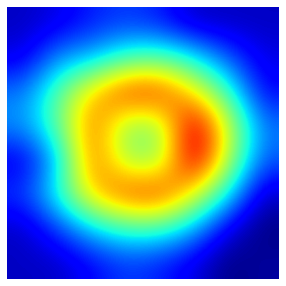} &
\includegraphics[width=0.13\textwidth]{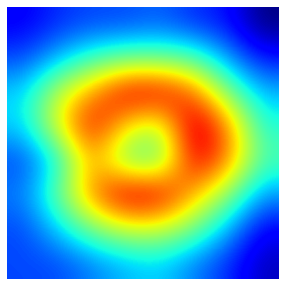} &
\includegraphics[width=0.13\textwidth]{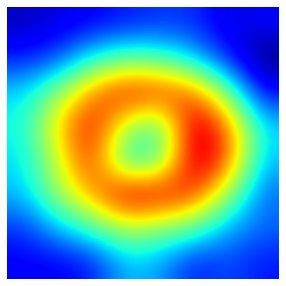} &
\includegraphics[width=0.13\textwidth]{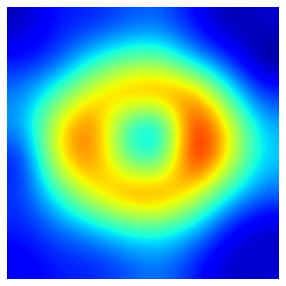} &
\includegraphics[width=0.13\textwidth]{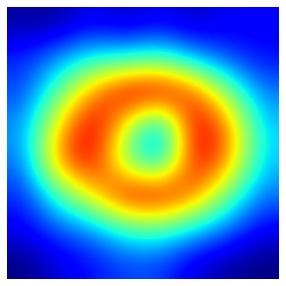} \\
R & \includegraphics[width=0.13\textwidth]{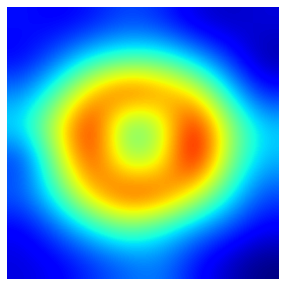} &
\includegraphics[width=0.13\textwidth]{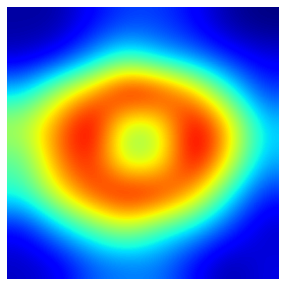} &
\includegraphics[width=0.13\textwidth]{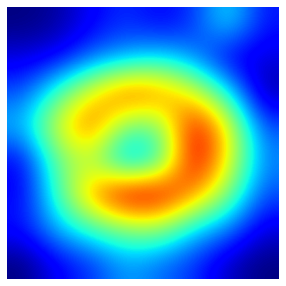} &
\includegraphics[width=0.13\textwidth]{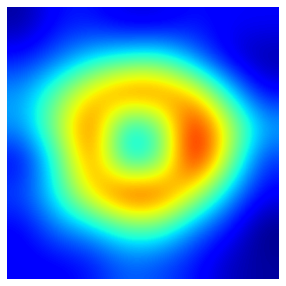} &
\includegraphics[width=0.13\textwidth]{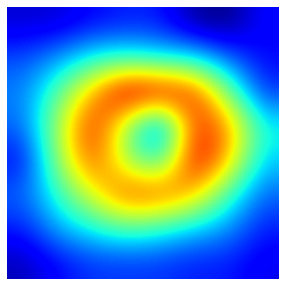} \\
\end{tabular}
\\
\vspace{0.25 cm}
MS samples, RNFL layer \\
\vspace{0.25 cm}
\begin{tabular}{cccccc}
& subject 1 & subject 2 & subject 3 & subject 4 & subject 5 \\
L & \includegraphics[width=0.13\textwidth]{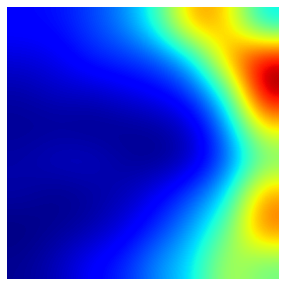} &
\includegraphics[width=0.13\textwidth]{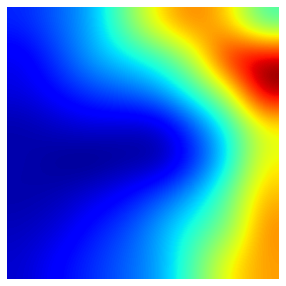} &
\includegraphics[width=0.13\textwidth]{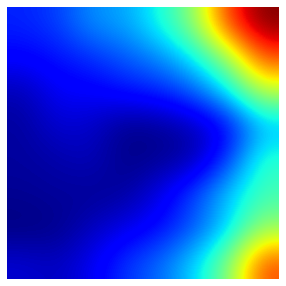} &
\includegraphics[width=0.13\textwidth]{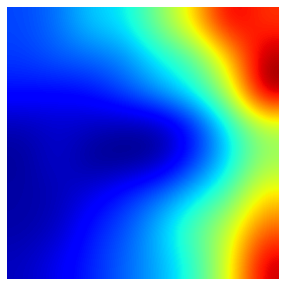} &
\includegraphics[width=0.13\textwidth]{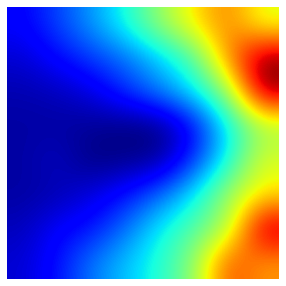} \\
R & \includegraphics[width=0.13\textwidth]{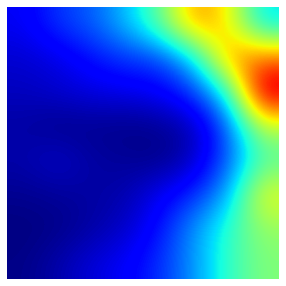} &
\includegraphics[width=0.13\textwidth]{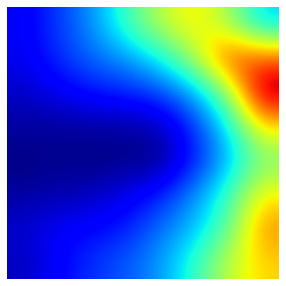} &
\includegraphics[width=0.13\textwidth]{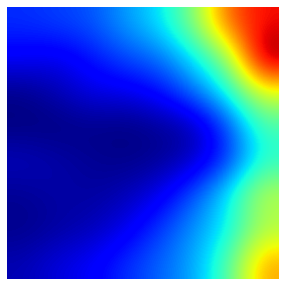} &
\includegraphics[width=0.13\textwidth]{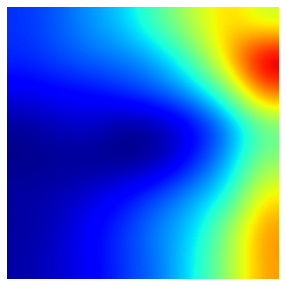} &
\includegraphics[width=0.13\textwidth]{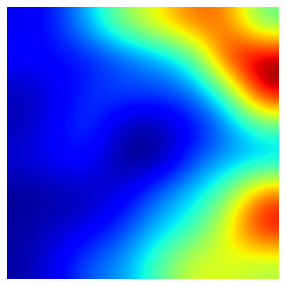} \\
\end{tabular}
\\
\vspace{0.25 cm}
HC samples, RNFL layer \\
\vspace{0.25 cm}
\begin{tabular}{cccccc}
& subject 6 & subject 7 & subject 8 & subject 9 & subject 10 \\
L & \includegraphics[width=0.13\textwidth]{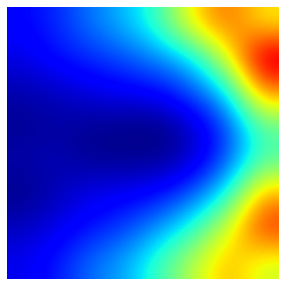} &
\includegraphics[width=0.13\textwidth]{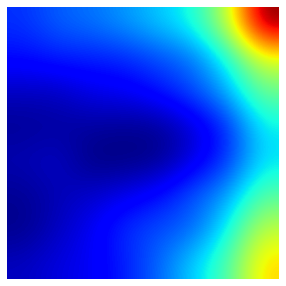} &
\includegraphics[width=0.13\textwidth]{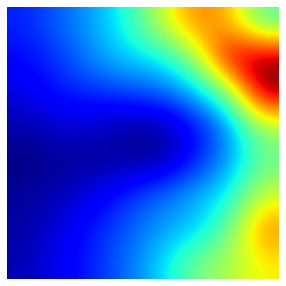} &
\includegraphics[width=0.13\textwidth]{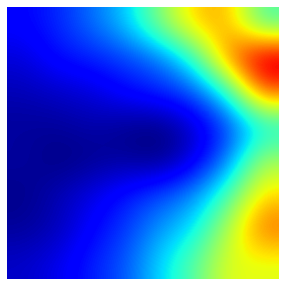} &
\includegraphics[width=0.13\textwidth]{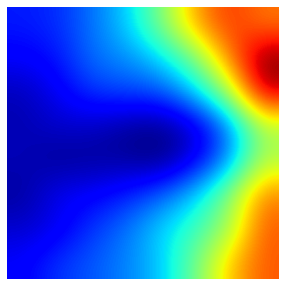} \\
R & \includegraphics[width=0.13\textwidth]{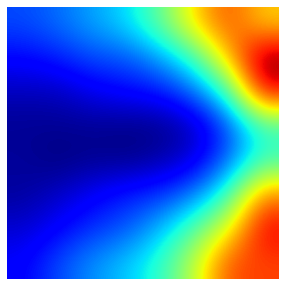} &
\includegraphics[width=0.13\textwidth]{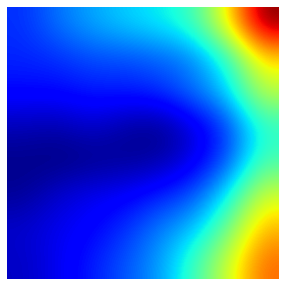} &
\includegraphics[width=0.13\textwidth]{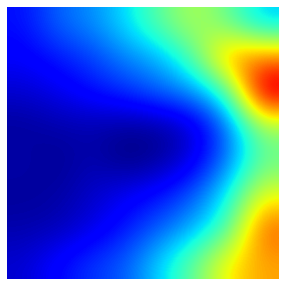} &
\includegraphics[width=0.13\textwidth]{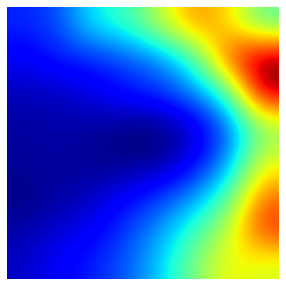} &
\includegraphics[width=0.13\textwidth]{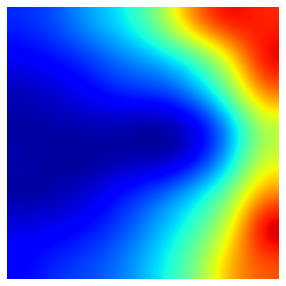} \\
\end{tabular}

\caption{Selection of illustrative samples of the Posterior pole (PPole) grid values found in our database.
The first row in each group shows the data acquired from the left eye while the
second row shows the data from the right eye.
Bicubic interpolation is used for enhancing visualization.}
\label{fig:Samples}
\end{figure}

Low-quality images (quality score higher than 25/40) or images with movement artifacts were excluded from the analysis.
The OCT device performed a detailed segmentation of the retinal layers that was used for obtaining the PPole measurements
as follows.
The $25^\textnormal{o} \times 30^\textnormal{o}$ area was divided into an 8 $\times$ 8 grid of $0.86 \times 0.86$ mm cells,
and the average thickness of each segmented layer was recorded in the 64 cells.
In addition, the 64 cells were grouped into 6 zones previously proposed in~\cite{Vilades_23} for a better clinical understanding
of the information.
The criteria for zone grouping clusters the upper (Zones 4 and 5), lower (Zones 3 and 6), and papillomacular bundle (Zone 1) arch patterns
that follows the RNFL, the concentric pattern of the GCL dividing quadrants (Zones 3, 4, 5, and 6),
and the macular area, with the highest density of ganglion cells (Zone 2).
Both the 64-dimensional and the 6-dimensional measurements will be considered feature sets in our work.
Figure~\ref{fig:PPoleZones} illustrates the division of the PPole grid into zones.
In this work, we will focus on the measurements of the ganglion cell layer (GCL) and the retinal nerve fiber layer (RNFL).
Table~\ref{table:Demographics} shows whether statistical significance can be obtained for the zone-based measurements, following
the study in~\cite{Martucci_21,Vilades_23}.
Figure~\ref{fig:Samples} shows typical examples of the PPole data for multiple sclerosis (MS) and healthy control (HC) records
in our database.

Visual inspection of the whole database allowed us to obtain initial insights into the group differences that may be interesting
for the diagnosis:
\begin{itemize}
\item There are visual differences between left and right eyes (e.g. subject 5 from MS group and subject 8 from HC group).
\item The GCL layer shows a torus-like shape that is frequently broken in the temporal-inferior region for the MS group (e.g. subject 1).
\item We can find this GCL break in the control group although to a lesser extent (e.g. subject 8).
\item The GCL torus thickness is usually greater in the HC group, although there are subjects with a thin thickness (e.g. subject 9).
\item The RNFL layer shows a wider dark blue extension for the MS subjects, which means that the thickness in the MS group tends to be
smaller than in the HC group (e.g. visual comparison between subject 3 and subject 7).
\end{itemize}

\section{Methods}
\label{sec:Methods}

\subsection{Machine learning methods considered in this work}

Our problem can be classified as a very small dataset problem.
The Zones feature set can be represented as tabular data.
Although the $8 \times 8$ PPole grid feature set could be represented as imaging data,
the low resolution of the images made us lean towards the tabular representation.
We included in our study two black-box methods, gradient boosting and random forest,
ranking as the best-performing methods in different applications and challenges~\cite{Marinescu_20}.
In addition, we included one good-performing glass-box method, explainable boosting machine,
in order to assess the accuracy-explainability trade-off for our application.
We discarded the use of neural networks architectures from our study since
the superiority of traditional methods in datasets with similar characteristics to ours
has been extensively demonstrated~\cite{Grinsztajn_22}.

\subsubsection{Gradient Boosting (XGB)}

Gradient Boosting is an ensemble learning technique that combines several weak learners (with poor accuracy)
into a strong learner (with high accuracy).
In particular, gradient boosting is a model ensemble of individual decision trees that are trained iteratively
such that each new tree improves the residual errors of the previous tree ensemble.
XGBoost (XGB) is an optimized distributed gradient boosting library available in the popular {\it scikit-learn}
library.
Its high efficiency is achieved through a parallel tree boosting algorithm that is known to accurately solve data
science problems involving billions of examples.
In the last years, XGB is ranked among the best methods for ML classification and regression on tabular data~\cite{Grinsztajn_22}.
To the date, neural networks are not able to outperform XGB in many different tasks.
For example, the winner of the Alzheimer's Disease diagnosis forecast in the TADPOLE Challenge was Frog, a system
built on a gradient boosting machine with XGBoost~\cite{Marinescu_20}.

The most relevant hyperparameters of gradient boosting are the learning rate (learning rate), maximum tree depth
(max depth), number of trees to fit (n estimators), and L2 regularization weight (reg lambda). We performed
hyperparameter selection using a Bayesian search with {\it optuna} package (\url{https://optuna.org/}).
We found that the hyperparameter selection greatly improved the validation accuracy but the default hyperparameters
provided a system whose test accuracy is close to that of the best-performing models in validation.
For the sake of reproducibility, XGBoost has been executed in our study with the default hyperparameter values:
learning rate=0.3, max depth=6, n estimators=100, and reg lambda=1.

\subsubsection{Random Forests (RF)}

Random Forests (RF) is a model ensemble that consists of a large number of decision trees.
Each individual tree in the forest performs one class prediction, and the class with the majority of the votes becomes
the model prediction. The trees in a forest are uncorrelated models that together produce ensemble predictions that are
more accurate than any of the individual predictions. Uncorrelation is the key to a successful ensemble. It is obtained
from random selection of different sets of data and different features for each tree.
RF has been proposed in different applications for building computer-aided diagnosis systems~\cite{ElSappagh_21,Fabrizio_21}.
The runner-up of the diagnosis forecast in the TADPOLE Challenge was ThreeDays, a random forest (RF) machine.

The most relevant hyperparameters of RF are the method for sampling data points (bootstrap), maximum number of levels in
the tree (max depth), number of features to be considered at every split (max features), minimum number of samples required
at each leaf (min samples leaf) and to split a node (min samples split), and number of trees in the forest (n estimators).
As with XGB, we performed hyperparameter selection using optuna. We also found that the default hyperparameters
provided the same accuracy as those of the best-performing model. Therefore, RF has
been executed in our study with the default hyperparameter values: bootstrap =
True, max depth = None, max features = auto, min samples leaf = 1,
min samples split = 2, n estimators = 100.

\subsubsection{Explainable Boosting Machine (EBM)}

Explainable Boosting Machine (EBM) is a glassbox model.
It was designed with the idea of obtaining accuracies comparable to state-of-the-art ML methods like boosted trees and random forests,
while being highly intelligible and explainable.
EBM fits under the family of Generalized Additive Models (GAM)

\begin{equation}
\label{eq:EBM}
 g(E[y]) = \beta_0 + \sum f_j(x_j),
\end{equation}

\noindent where the feature functions $f_j$ are estimated using small trees in a way that it resembles gradient boosting.
The boosting procedure is restricted to train one feature at a time in all vs all fashion using a very low learning rate so that feature
order does not matter.
Thus, the feature functions are composed of sums of small trees.
Therefore, many complicated functions can be modeled very accurately due to the versatility of trees.
In addition, this way of training mitigates the effects of co-linearity.
The best feature functions $f_j$ provide measurements of how each feature contributes to the model's prediction of the problem yielding
the explainability of the model.
Indeed, EBM can automatically detect and include pairwise interaction terms by using models of the form

\begin{equation}
\label{eq:EBMi}
 g(E[y]) = \beta_0 + \sum f_j(x_j) + \sum f_{i,j}(x_i, x_j)
\end{equation}

\noindent which further increases the accuracy while maintaining explainability in an analogous way.
In this work, we will refer with EBM to the model given by Equation~\ref{eq:EBM} and with EBM + i to the model with interactions
given by Equation~\ref{eq:EBMi}.



In general, the default hyperparameters for EBM should perform reasonably well on most problems.
The authors recomend in~\url{https://interpret.ml/docs/faq.html} to train a model with the defaults and use explainability
to catch features with abnormal behavior.
Using the defaults, we obtained a performance comparable with XGB and RF and we did not detected any problem requiring
hyperparameter tuning.

\section{Explainable machine learning}
\label{sec:xAI}

Explainable AI (xAI) is a recent subfield of AI that aims to provide explanations of general machine learning algorithms.
xAI methods are intended to be combined with the most complex families of methods that have been traditionally considered
as black boxes~\cite{Adadi_17,Molnar_2021}.
The objective of xAI is to overcome the trade-off traditionaly existing between accuracy and explainability and provide
both powerful and explainable systems with arguments for increasing the confidence in the output of the algorithms.

Explainability is an important first step toward fully usable and trustworthy AI systems.
The most important current xAI techniques facilitate the establishment of the importance given to the different features by
the system, assessing how a particular feature affects model predictions, or which feature values favor or hinder correct
or incorrect predictions.
All this information can help ensure whether only significant features coherent with current knowledge are used to infer
the output for the application of interest, or, on the other hand, to assess whether the system is biased or
the experimental design is flawed.

SHapley Additive exPlanations (SHAP) has attracted increasing attention in the field of xAI (https://shap.readthedocs.io).
The core idea of SHAP is to transfer ideas from cooperative game theory to the attribute feature importance of a model output
given an input~\cite{Lundberg_17}.
SHAP values represent the distribution of the contributions toward game success or failure amongst all the players working
in cooperation.

In the context of explainability of machine learning methods, SHAP values represent the change in each feature in the expected
model prediction under conditioning on that feature.
The explainability from SHAP is achieved through different graphical representations derived from the SHAP values.
In this work, we will represent the mean of the SHAP values as box plots, and represent the impact of feature values on
the probability of the MS class using violin plots and partial dependence plots.
In addition, SHAP allows providing local explanations on the decisions taken by the models.
In this work, we will use waterfall plots.
In ~\url{https://www.kaggle.com/code/meliao/shap-on-titanic-why-is-rose-alive-but-jack-not}
we can find an easy to follow example
on the use of SHAP and the graphical representation of the SHAP values for understanding the survival chances of Rose and Jack
in Titanic's tragedy.

In EBM, the contribution of each feature to a final prediction can be visualized and understood by plotting $f_j$ or $f_{i,j}$.
Since EBM is an additive modeling, each feature contributes to predictions in a modular way that makes it easy to
reason about the contribution of each feature to the prediction (similarly to SHAP).
For example, term contributions can be sorted and visualized to show which features have the most impact on any individual prediction.


\section{Experimental design}
\label{sec:Implementation}


Ideally, we would like to work with a train/validation/test split.
However, our data set is small.
Therefore, it is more convenient to evaluate the performance of our methods and study the explainability
of our models using ten-fold cross-validation splits.
Thus, the original set was divided into ten different train-test splits of subjects with $80 \% - 20 \%$
proportionality.
This means that if a subject falls into the train/test set of a given fold, the eyes for this subject belong
to the train/test set.
Since we were not separating a hold-out test from the very beginning of the process, we were extremely careful with
our experimental setup in order to avoid any kind of data leakage.

Since each subject can contribute to the dataset with one or two eyes, we generated four different datasets
and assessed the performance of the corresponding models: left eye (L), right eye (R), randomly selected eye (rand),
and both eyes (LR) (independent sample assumption).
For the randomly selected eye models, random selection was only performed for subjects with both eyes.
We carried out a comparison of the performance metrics and selected the most appropriate models for our interpretability
study, after making sure that there was not statistical significance with the other candidates.
We ended up with ten different models with different accuracies for assessing explainability.
We opted for showing the explainability results for the best-performing fold.
The Supplementary Material shows the explainability results for the worst-performing fold.

Although the PPole grid features have a two-dimensional structure, we decided to represent them as tabular data,
given the low resolution and the small number of samples. Future work will address the problem for 2D.


Both XGB, RF, and EBM require little data preparation in contrast to other methods like neural networks (e.g. no normalization).
In addition, the default parameters should perform reasonably well on most problems.
As we mentioned in Section~\ref{sec:Methods}, preliminary exploration with Bayesian search did not provide a better hyperparameter setting
than default values.
We found similar results in a previous study on the Alzheimer's Disease Neuroimaging Initiative (ADNI)~\cite{Hernandez_22}.
Since further dividing our train set into train and validation sets would aggravate the small sample problem,
and there would be a combinatorial explosion in the number of experiments, we decided to use the default hyperparameters.

Finally, we considered both undersampling and oversampling for dealing with the problem of data unbalancing.
Preliminary exploration indicated that oversampling using SMOTE on the minority group (MS) was preferable.
It should be noticed that the test set should be unbalanced, preserving the actual distribution of the overall population
of our dataset.
Otherwise, we have the risk of providing an overestimation of the true capacities of the models in the evaluation.

Our methods were implemented in Python 3.6 using the libraries in {\it scikit-learn} (https://scikit-learn.org).
For XGB and RF, SHAP values were obtained from tree explainers, a fast implementation of SHAP for tree-based methods.
Our experiments were run on a machine with an Intel Core i7 working at 3.70 GHz and 32 GBS of DDR3 RAM.


\section{Results}
\label{sec:Results}


\subsection{Performance study}

The performance in this study has been measured from the metrics typically used in ML studies that are computed from the
number of true positives, true negatives, false positives, and false negatives in the test set.
These metrics are the accuracy, sensitivity, specificity, F1-score, and area under the curve (AUC).
Figure~\ref{fig:BoxPlots} shows the distribution of the obtained results for the populations with left eyes (L), right eyes (R),
randomly selected eyes (rand), and both eyes (LR).
The Supplementary Material gathers the results obtained through the ten-fold cross-validation experiments.
From the analysis of the plots, we identify the most remarkable results:

\begin{figure}[!t]
\centering
\scriptsize
\begin{tabular}{cccc}
GCL, Zones & \hspace{-0.75 cm} GCL, PPole grid & \hspace{-0.75 cm} RNFL, Zones & \hspace{-0.75 cm} RNFL, PPole grid \\
\includegraphics[width=0.25\textwidth]{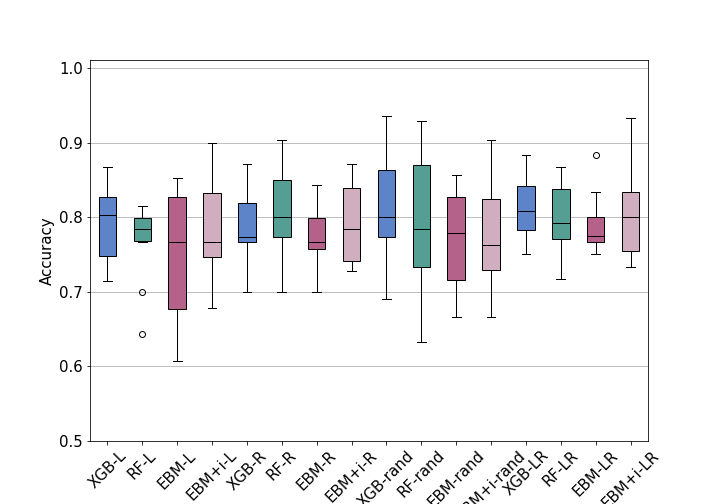} &
\hspace{-0.75 cm}
\includegraphics[width=0.25\textwidth]{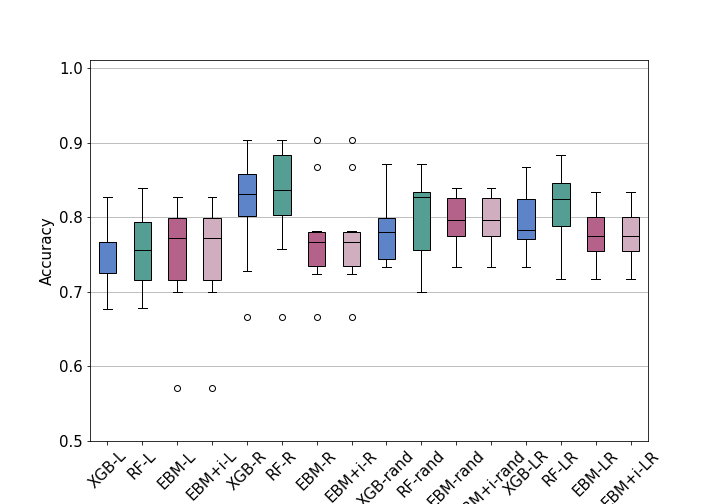} &
\hspace{-0.75 cm}
\includegraphics[width=0.25\textwidth]{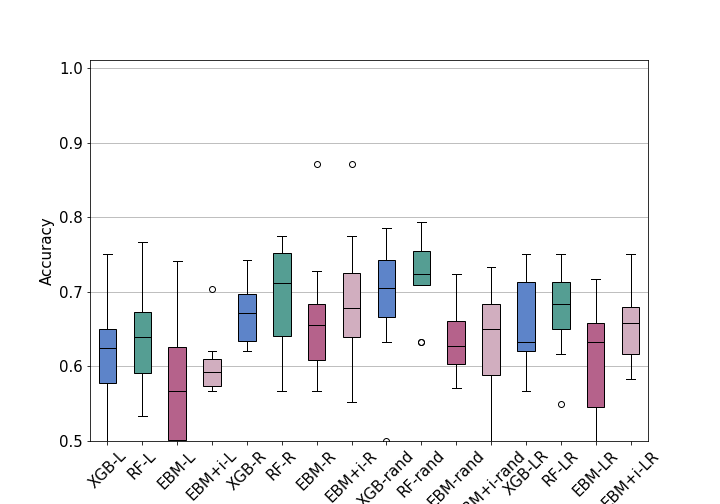} &
\hspace{-0.75 cm}
\includegraphics[width=0.25\textwidth]{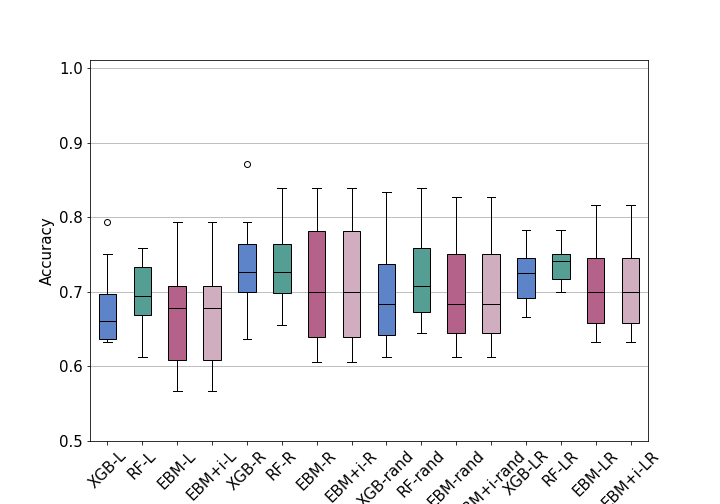} \\
\includegraphics[width=0.25\textwidth]{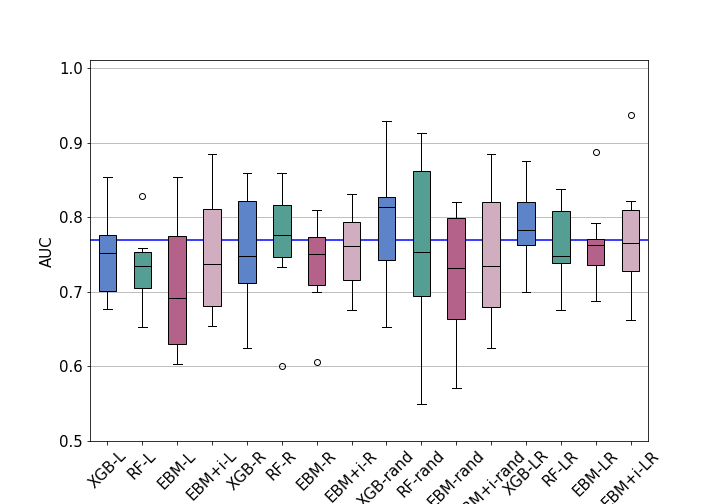} &
\hspace{-0.75 cm}
\includegraphics[width=0.25\textwidth]{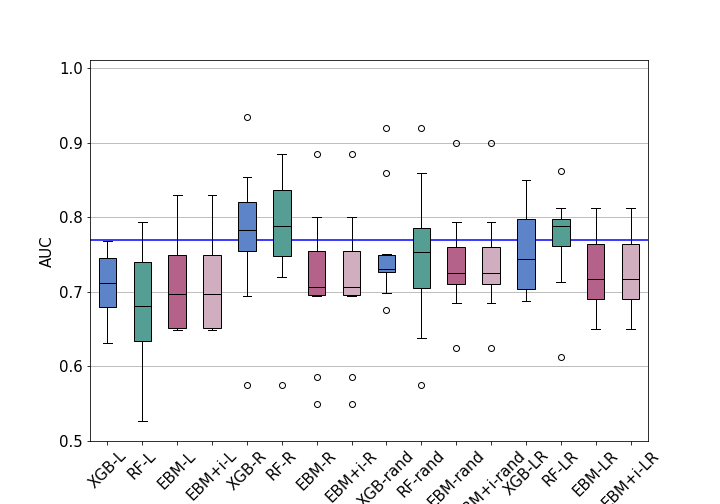} &
\hspace{-0.75 cm}
\includegraphics[width=0.25\textwidth]{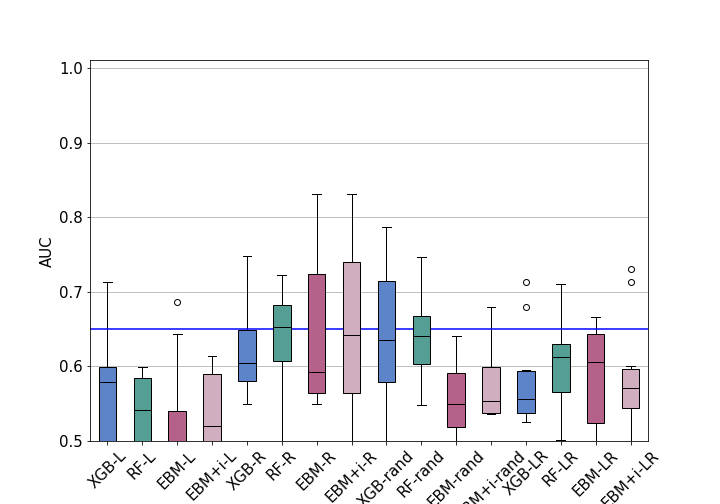} &
\hspace{-0.75 cm}
\includegraphics[width=0.25\textwidth]{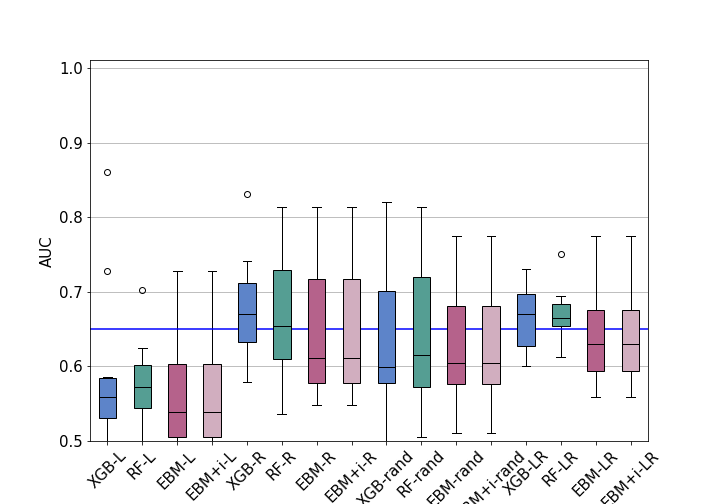} \\
\includegraphics[width=0.25\textwidth]{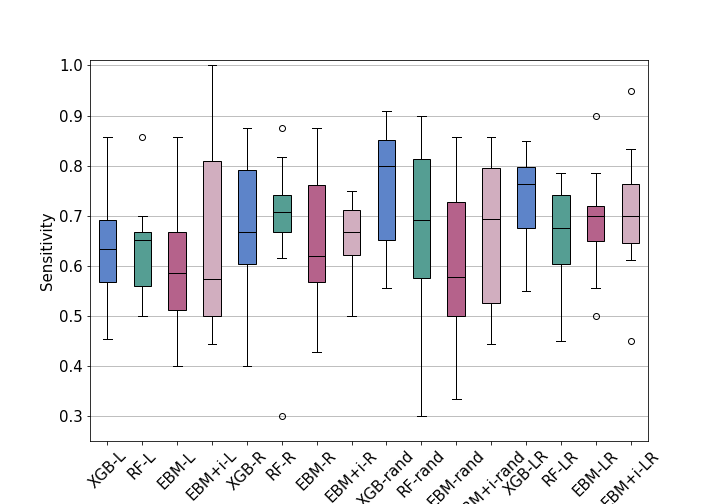} &
\hspace{-0.75 cm}
\includegraphics[width=0.25\textwidth]{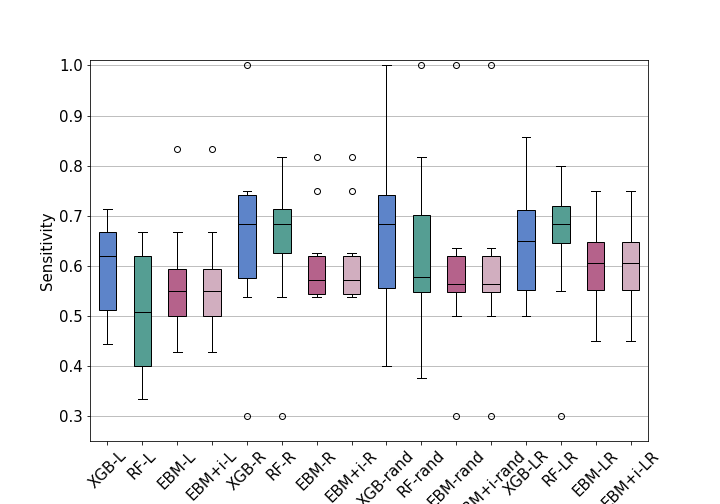} &
\hspace{-0.75 cm}
\includegraphics[width=0.25\textwidth]{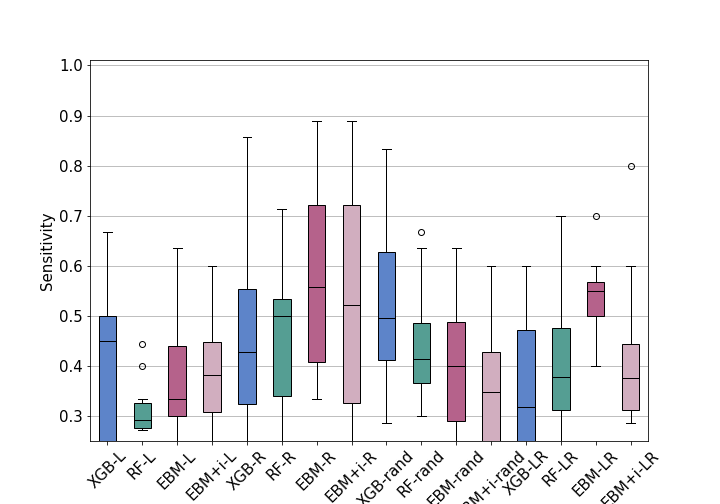} &
\hspace{-0.75 cm}
\includegraphics[width=0.25\textwidth]{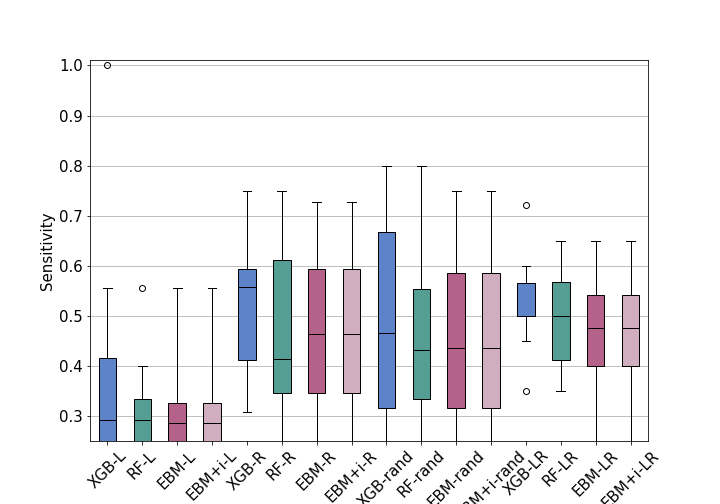} \\
\includegraphics[width=0.25\textwidth]{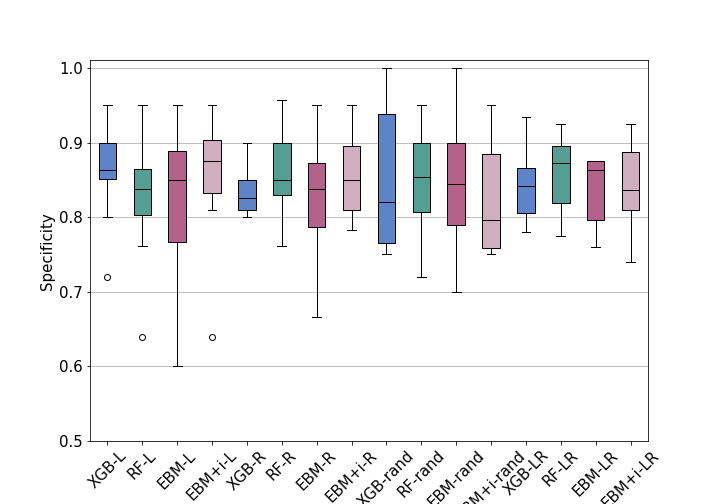} &
\hspace{-0.75 cm}
\includegraphics[width=0.25\textwidth]{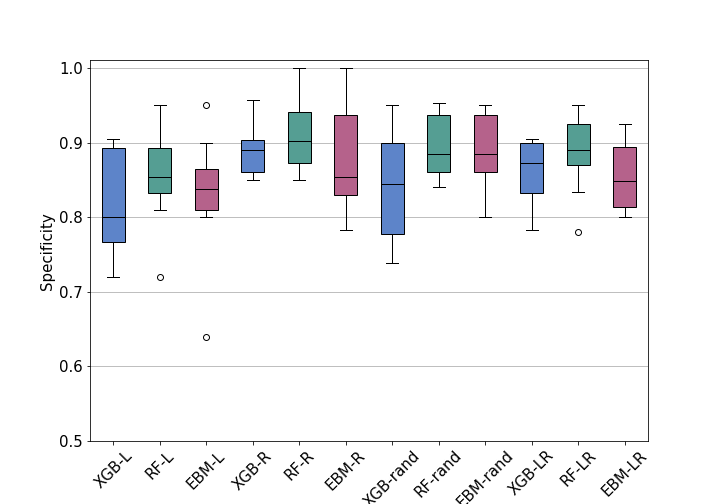} &
\hspace{-0.75 cm}
\includegraphics[width=0.25\textwidth]{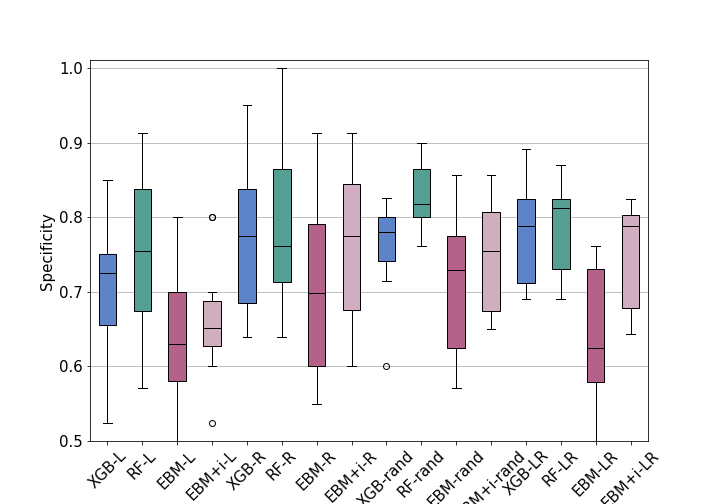} &
\hspace{-0.75 cm}
\includegraphics[width=0.25\textwidth]{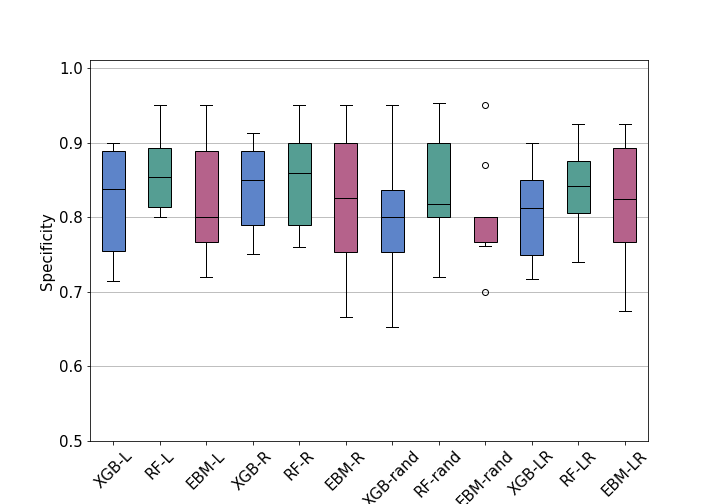} \\
\end{tabular}
\caption{Distribution of the performance metrics in the test sets of the ten-fold cross-validation experiments.
For each retinal layer, the left column shows the results obtained with the Zones feature set.
The right column shows the results obtained with the Posterior Pole (PPole) grid feature set.
L indicates the model built with the population of left eyes.
R indicates the model built with right eyes.
Rand indicates the model built with randomly selected eyes.
LR indicates the model with both eyes.
Blue boxes show the results obtained with gradient boosting (XGB), green with random forests (RF), and magenta with explainable boosting machine (EBM).
The horizontal line in the area under the curve (AUC) plot indicates the test set baseline provided in~\cite{Kenney_22}.
}
\label{fig:BoxPlots}
\end{figure}

\begin{itemize}
 \item The models built with the left eyes show in general a performance lower than the models built with the right eyes.
 \item The models built with randomly selected eyes show a performance similar to the models built with both eyes.
 \item The variance in accuracy is close to or greater than 5\% in the majority of models.
 \item The use of the GCL layer measurements as a feature set greatly outperforms the use of the RNFL layer.
 \item For the GCL layer,
 \begin{itemize}
 \item The Zones feature set provides a slightly improved performance with respect to the PPole grid feature set.
 \item The sensitivity indicates that some models obtain a low accuracy in the diagnosis of MS.
 This could be due to the small size of the dataset and the unbalanced nature of the sample.
 \item The specificity indicates that the models are in general good in the identification of HC.
 \item No method consistently stands out in performance above the rest.
 \end{itemize}
\end{itemize}


Welch's t-test on the GCL measurements and the Zones feature set did not find any statistical significance between
accuracies.
The same test on the PPole feature set showed statistical significance between L and R eyes for XGB (p = 0.02) and RF (p = 0.02),
L and rand eyes for RF (p = 0.07), EBM (p = 0.09), and EBM + i (p = 0.07),
and L and LR eyes for XGB (p = 0.03) and RF (p = 0.01).
Therefore, no statistical significance was found between rand and LR eyes.


Regarding the rule of thumb of having a number of samples per class of at least ten times the number of features,
we would like to remark that the Zones feature set meets this requirement. Although the PPole grid feature set
does not meet the requirement, it seems that the performance is not greatly affected by the ML methods used in
this study.
We observed similar behavior in our previous work for XGB and RF with the Alzheimer's Disease Neuroimaging Initiative
(ADNI) tabular data~\cite{Hernandez_22}.

Comparing our AUCs with the baseline provided in the test set in Kenney et al.~\cite{Kenney_22} for the average of GCL+ thickness from
both eyes, we can see that the rand models from the Zones feature set achieve a close mean. It is over the baseline for XGB.
LR models from the Zones feature set achieve a comparable mean in all cases.
The LR models from the PPole grid feature set slightly perform under with the exception of RF for the LR models.
For the RNFL thickness, our rand models for the Zones feature set show a comarable mean for XGB and RF.
The LR models from the Zones feature set underperforms the baseline.
All the LR models from the PPole grid feature set show a comparable mean.
It should be noticed that Kenney et al. used random sampling and the pRNFL thickness instead of RNFL.

Table~\ref{table:BestPerforming} shows the results of the best-performing random and LR models for each feature set configuration.
For the GCL layer, the models built with the random dataset greatly outperformed the models built with the LR dataset for
the Zones feature set in XGB and RF.
The remaining models performed similarly regardless of the use of random or LR datasets.
Therefore, in the following, we proceed with the interpretability study for the models built with the LR dataset
since the performance is similar and it may give us more opportunities to analyze interesting cases.
Furthermore, the fold identifier (id) of the best-performing model is the same for all three methods, and therefore the results can
be compared fairly.
With the exception of Subsection~\ref{subsec:Grids}, our interpretability will focus on assessing the measurements obtained on the GCL layer.

\begin{table}[!t]
\begin{center}
\scriptsize
Random
\begin{tabular}{|c|c|c|c|c|c|c|c|c|}
\hline
Method & Layer & Feature set & fold id & Acc (\%) & Sens (\%) & Spec (\%) & F1-score (\%) & AUC \\
\hline
XGB & GCL & Zones & 5 & 93.54 & 90.90 &	95.00 & 90.90 &	0.92 \\
XGB & GCL & PPole grid & 5 & 87.09 & 81.81 & 90.00 & 81.81 & 0.85 \\
RF & GCL & Zones & 1 & 92.85 & 87.50 & 95.00 & 87.50 & 0.91 \\
RF & GCL & PPole grid & 5 & 87.09 & 81.81 &	90.00 & 81.81 & 0.85 \\
EBM & GCL & Zones & 1 & 85.71 & 60.00 & 100.00 & 66.66 & 0.75 \\
EBM & GCL & PPole grid & 4 & 83.33 & 100.00 & 80.00 & 66.66 & 0.90 \\
EBM + i & GCL & Zones & 5 & 90.32 &	81.81 &	95.00 & 85.71 &	0.88 \\
EBM + i & GCL & PPole grid & 5 & 83.87 & 81.81 & 85.00 & 78.26 & 0.83 \\
\hline
XGB & RNFL & Zones & 1 & 78.57 & 75.00 & 80.00 & 66.66 & 0.77 \\
XGB & RNFL & PPole grid & 5 & 77.41 & 72.72 & 80.00 & 69.56 & 0.76 \\
RF & RNFL & Zones & 0 & 79.31 & 66.66 & 82.60 & 57.14 &	0.74 \\
RF & RNFL & PPole grid & 5 & 83.87 & 72.72 & 90.00 & 76.19 & 0.81 \\
EBM & RNFL & Zones & 0 & 72.41 & 50.00 & 78.26 & 42.85 & 0.64 \\
EBM & RNFL & PPole grid & 0 & 82.75 & 66.66 & 86.95 & 61.53 & 0.76 \\
EBM + i & RNFL & Zones & 4 & 73.33 & 60.00 & 76.00 & 42.85 & 0.68 \\
EBM + i & RNFL & PPole grid & 4 & 76.66 & 80.00 & 76.00 & 53.33 & 0.78 \\
\hline
\end{tabular}
LR
\begin{tabular}{|c|c|c|c|c|c|c|c|c|}
\hline
Method & Layer & Feature set & fold id & Acc (\%) & Sens (\%) & Spec (\%) & F1-score (\%) & AUC \\
\hline
XGB & GCL & Zones & 6 & 85.00 & 78.57 & 86.95 & 70.96 &	0.82 \\
XGB & GCL & PPole grid & 5 & 86.66 & 80.00 & 90.00 & 80.00 & 0.85 \\
RF & GCL & Zones & 5 & 86.66 & 75.00 & 92.50 & 78.94 & 0.83 \\
RF & GCL & PPole grid & 5 & 88.33 &	80.00 & 92.50 & 82.05 & 0.86 \\
EBM & GCL & Zones & 5 & 88.33 & 90.00 & 87.50 &  83.72 & 0.88 \\
EBM & GCL & PPole grid & 5 & 83.33 & 75.00 & 87.50 & 75.00 & 0.81 \\
EBM + i & GCL & Zones & 5 & 93.33 & 95.00 & 92.50 & 90.47 & 0.93 \\
EBM + i & GCL & PPole grid & 5 & 86.66 & 70.00 & 95.00 & 77.77 & 0.82 \\
\hline
XGB & RNFL & Zones & 5 & 75.00 & 28.57 & 89.13 & 34.78 & 0.58 \\
XGB & RNFL & PPole grid & 5 & 78.33	& 55.00 & 90.00 & 62.85 & 0.72 \\
RF & RNFL & Zones & 0 & 75.00 & 35.71 &	86.95 &	40.00 & 0.61 \\
RF & RNFL & PPole grid & 5 & 78.33 & 65.00 & 85.00 & 66.66 & 0.75 \\
EBM & RNFL & Zones & 0 & 71.66 & 57.14 & 76.08 & 48.48 & 0.66 \\
EBM & RNFL & PPole grid & 5 & 81.66 & 65.00 & 90.00 & 70.27 & 0.77 \\
EBM +  i & RNFL & Zones & 4 & 68.33 & 80.00 & 66.00 & 45.71 & 0.73 \\
EBM + i & RNFL & PPole grid & 1 & 76.66 & 60.00 & 85.00 & 63.15 & 0.72 \\
\hline
\end{tabular}
\end{center}
\caption{
Results of the best-performing random and LR (both eyes) models for each feature set configuration considered
in this work.
We indicate the id of the training and test set used to generate the model and obtain the metrics.
}
\label{table:BestPerforming}
\end{table}


%


\subsection{Explainability study}

\subsubsection{Global SHAP-based explainability}

\begin{figure}[!t]
\centering
\scriptsize
XGB \\
\begin{tabular}{ccc}
\includegraphics[width=0.3\textwidth]{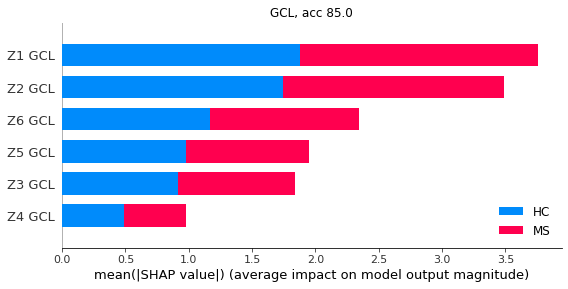} &
\includegraphics[width=0.3\textwidth]{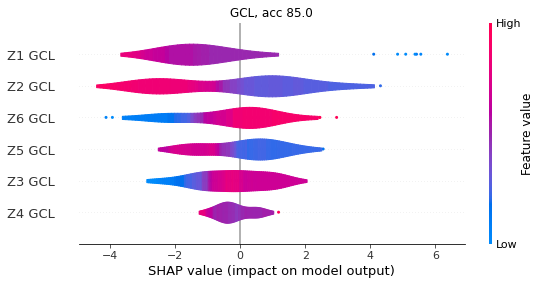} &
\includegraphics[width=0.3\textwidth]{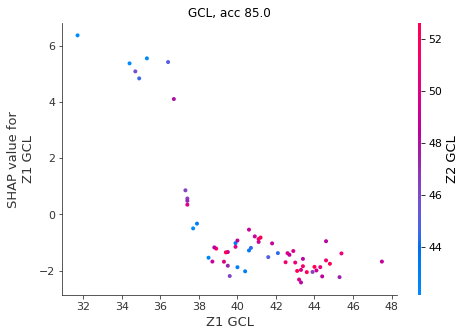}
\\
\end{tabular}
RF \\
\begin{tabular}{ccc}
\includegraphics[width=0.3\textwidth]{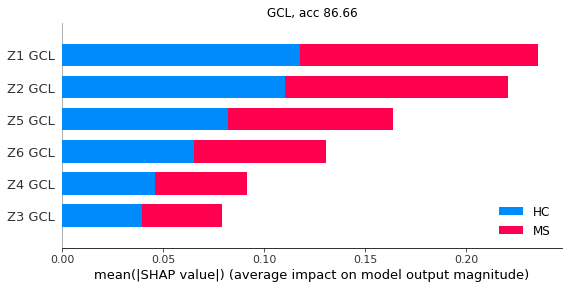} &
\includegraphics[width=0.3\textwidth]{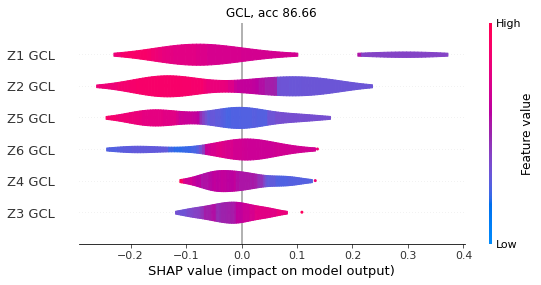} &
\includegraphics[width=0.3\textwidth]{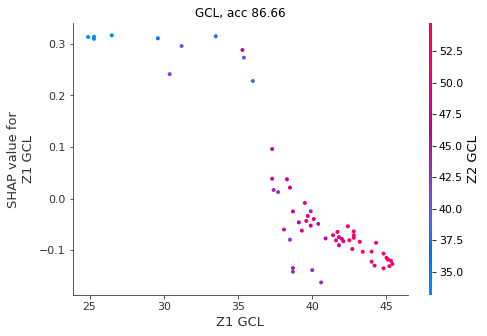}
\\
\end{tabular}
\caption{{Global SHAP results.} SHAP results for the gradient boosting (XGB) and random forests (RF) best models obtained with the Zones feature set.
Left, horizontal bar plots of the SHAP values.
Center, violin plots.
Right, partial dependence plot for the two most important features.}
\label{fig:SHAPZonesBest}
\end{figure}

Figure~\ref{fig:SHAPZonesBest} shows different representations of the average SHAP values for the best-performing
XGB and RF models with the Zones feature set. These representations include horizontal bar plots of the SHAP values ranked by order of importance,
violin plots with the distribution of the SHAP values colored by the magnitude of the feature values,
and
partial dependence plots for assessing the marginal effect of the two most important features on the predicted outcome of the ML models.


For both XGB and RF, zones Z1 and Z2 are the most relevant for the discrimination of MS from HC.
The order of importance is similar.
The violin plots show that medium and low values of the average thicknesses contribute to a
higher probability for MS.
The only exception is for Z6 in the case of XGB, where high values of the average thickness
contribute to a higher probability for MS.
Our clinical experts indicated that this zone corresponds with the most inferior and temporal paramacular area.
This is a zone with medium and large size blood vessels (see Figure 2) and, therefore, GCL measurements
may be less accurate.
The partial dependence plots show that the slope of the distribution of the SHAP values for
Z1 is negative. This means that high values of the average thickness contribute to a lower
probability for MS, corroborating the violin plot for Z1.
The same happens with Z2.

\begin{figure}[!htpb]
\centering
\scriptsize
XGB \\
\begin{tabular}{cc}
\includegraphics[width=0.5\textwidth]{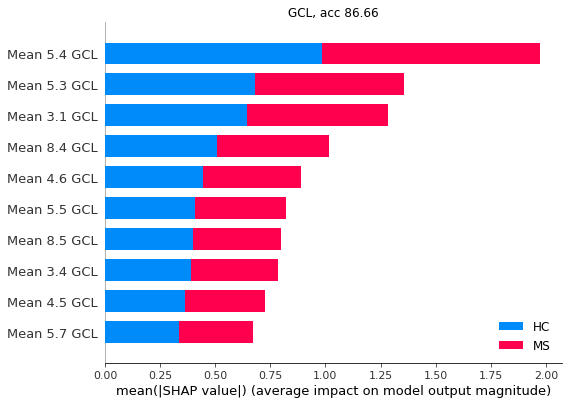} &
\includegraphics[width=0.5\textwidth]{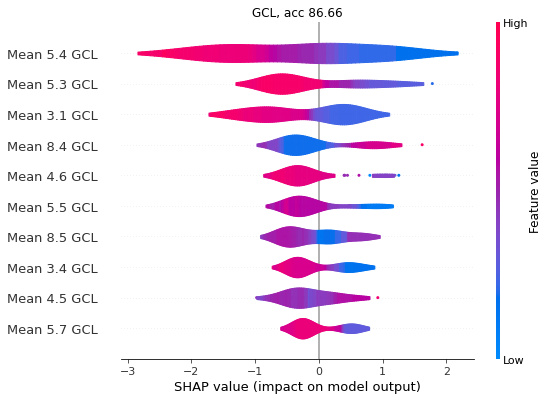}
\\
\includegraphics[width=0.5\textwidth]{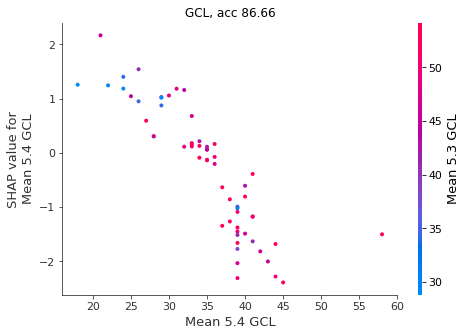} &
\includegraphics[width=0.5\textwidth]{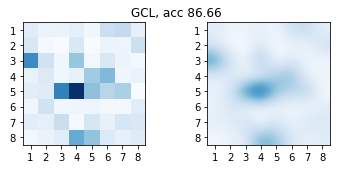}
\\
\end{tabular}
RF \\
\begin{tabular}{cc}
\includegraphics[width=0.5\textwidth]{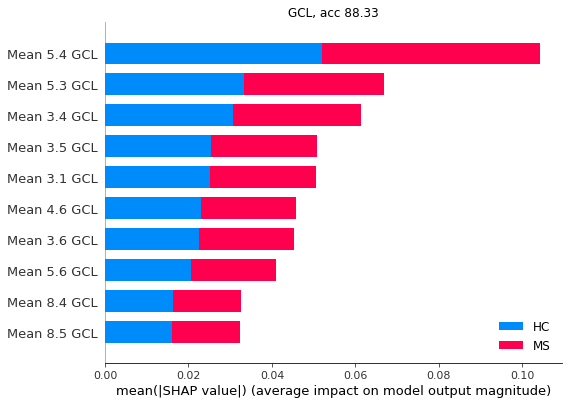} &
\includegraphics[width=0.5\textwidth]{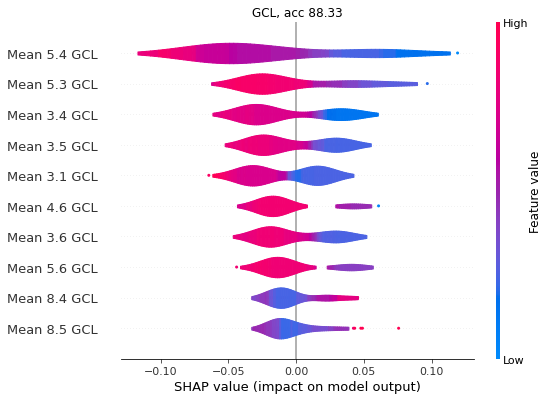}
\\
\includegraphics[width=0.5\textwidth]{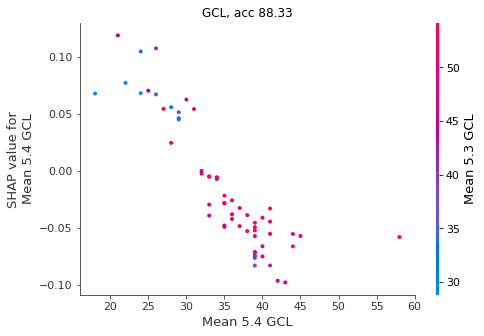} &
\includegraphics[width=0.5\textwidth]{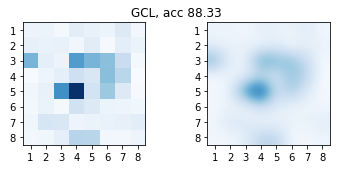}
\\
\end{tabular}
\caption{Global SHAP results. SHAP values for the gradient boosting (XGB) and random forests (RF) best models obtained with the posterior
pole (PPole) grid feature set.
The grid showing the SHAP values obtained in each spatial location is shown in its original resolution (left) and
after bicubic interpolation (right).}
\label{fig:SHAPPPoleBest}
\end{figure}

Analogously for PPole grid features, Figure~\ref{fig:SHAPPPoleBest} shows different representations
of the average 
of the SHAP values for the best-performing XGB and RF models. In addition to the bar, violin, and dependence plots,
we show the
SHAP values on the 8 $\times$ 8 grid for a finer location of the anatomical part identified with our
models as mostly affected by the disease.


For both XGB and RF, features 5.4 and 5.3 are the most relevant for the discrimination of MS from HC.
These features are located at Z1 and Z2 zones, which were marked by SHAP as the most important ones in the corresponding models (Figure~\ref{fig:SHAPZonesBest}).
The violin plots show that low average thicknesses contribute to a higher probability for MS.
Some exceptions to the rule can be identified for features located at the boundary points (e.g. feature 8.4).
The partial dependence plots show a negative slope corroborating the relationship between high thickness values and
low probability values for MS given by the models.
The grids show that RF feature importance is distributed among features 5.4 and 5.3 and some features located over the
superior and nasal superior locations. Therefore, RF seems to realize the 2D geometry of the data.
It is worth noticing that feature 3.1 is considered as important by both methods.
Our clinical experts found a plausible interesting interpretation of this finding that is discussed in Section~\ref{sec:Discussion}.
However, we do not rule out either a bias of our models due to the characteristics of our dataset.


\subsubsection{Local SHAP-based explainability}

\begin{figure}[!htpb]
\centering
\scriptsize
\begin{tabular}{cccc}
Best XGB & Best RF & Best EBM & Best EBM + i \\
\includegraphics[width=0.22\textwidth]{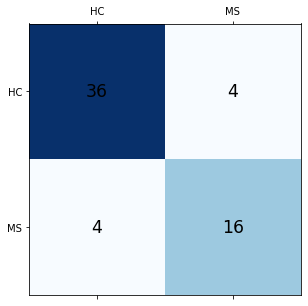} &
\includegraphics[width=0.22\textwidth]{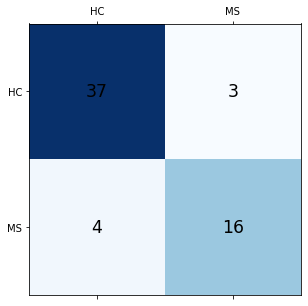} &
\includegraphics[width=0.22\textwidth]{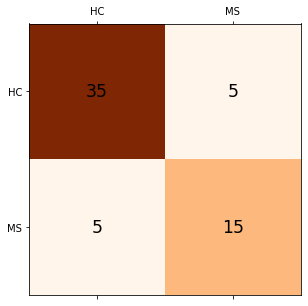} &
\includegraphics[width=0.22\textwidth]{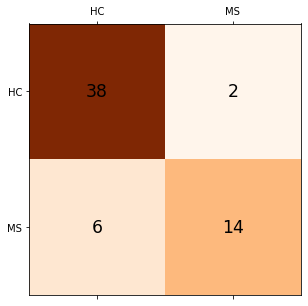}
\\
\end{tabular}
\caption{Confusion matrices for the best gradient boosting (XGB), random forests (RF), explainable boosting machine (EBM), and
EBM with interactions (EBM + i) models with the PPole grid feature set.}
\label{fig:ConfusionMatrices}
\end{figure}

\begin{figure}[!htpb]
\centering
\scriptsize
XGB \\
\begin{tabular}{ccc}
\includegraphics[width=0.15\textwidth]{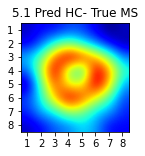} &
\includegraphics[width=0.3\textwidth]{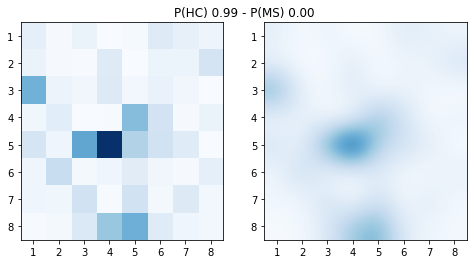} &
\includegraphics[width=0.3\textwidth]{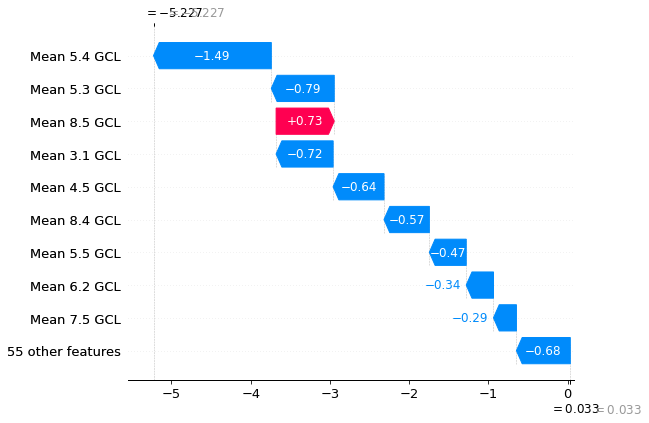} \\
\includegraphics[width=0.15\textwidth]{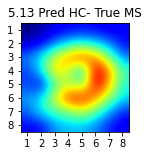} &
\includegraphics[width=0.3\textwidth]{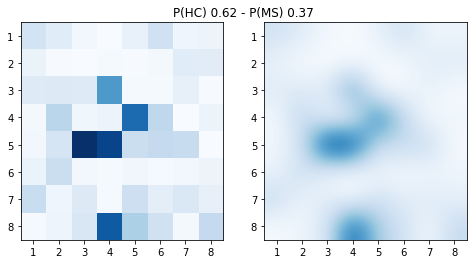} &
\includegraphics[width=0.3\textwidth]{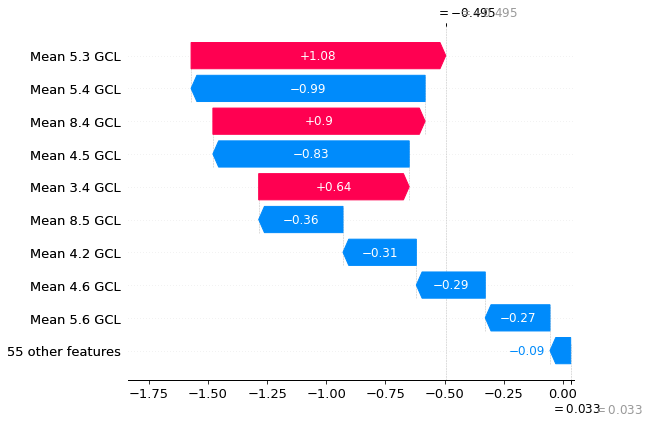} \\
\includegraphics[width=0.15\textwidth]{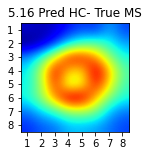} &
\includegraphics[width=0.3\textwidth]{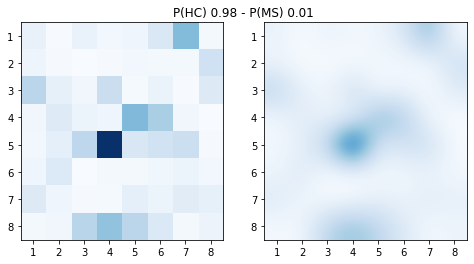} &
\includegraphics[width=0.3\textwidth]{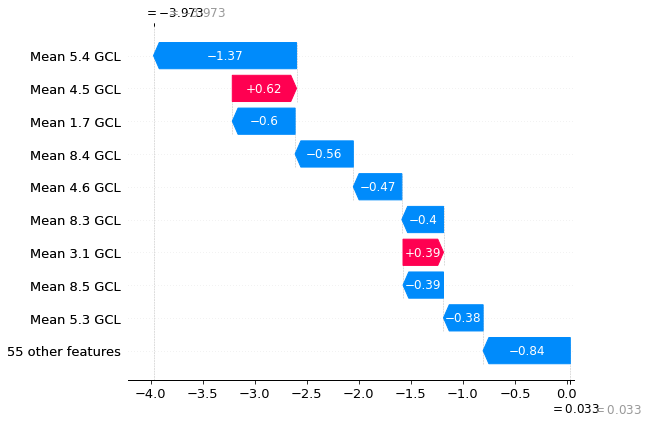} \\
\includegraphics[width=0.15\textwidth]{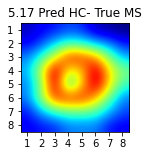} &
\includegraphics[width=0.3\textwidth]{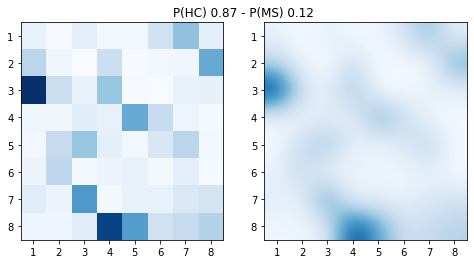} &
\includegraphics[width=0.3\textwidth]{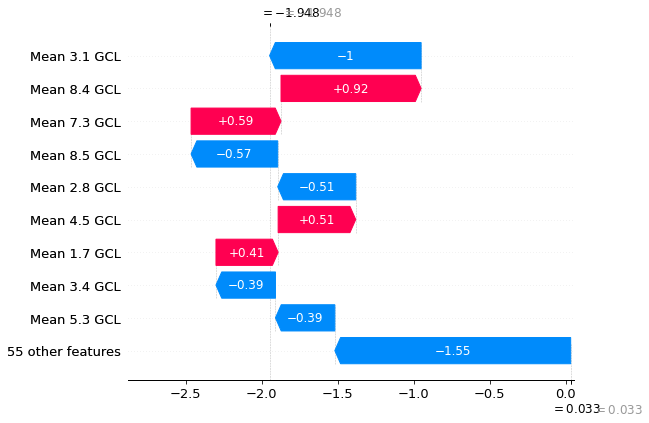} \\
\includegraphics[width=0.15\textwidth]{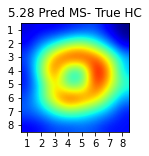} &
\includegraphics[width=0.3\textwidth]{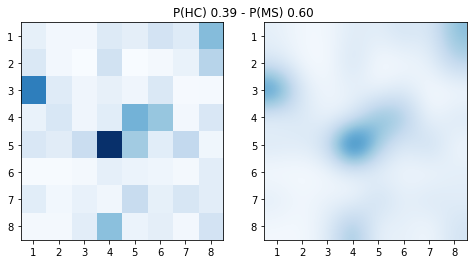} &
\includegraphics[width=0.3\textwidth]{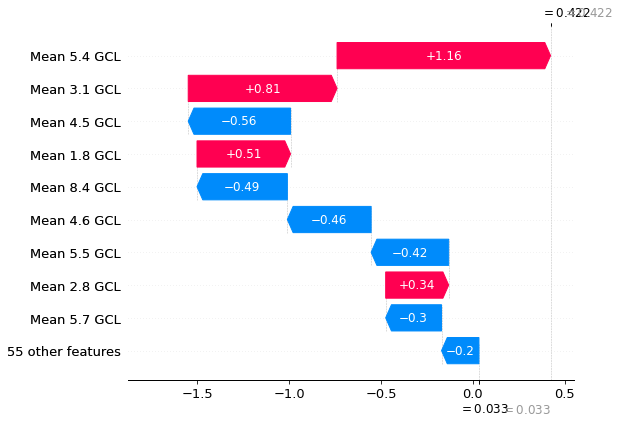} \\
\includegraphics[width=0.15\textwidth]{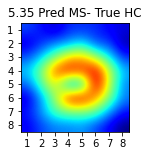} &
\includegraphics[width=0.3\textwidth]{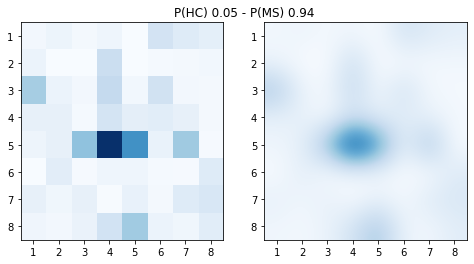} &
\includegraphics[width=0.3\textwidth]{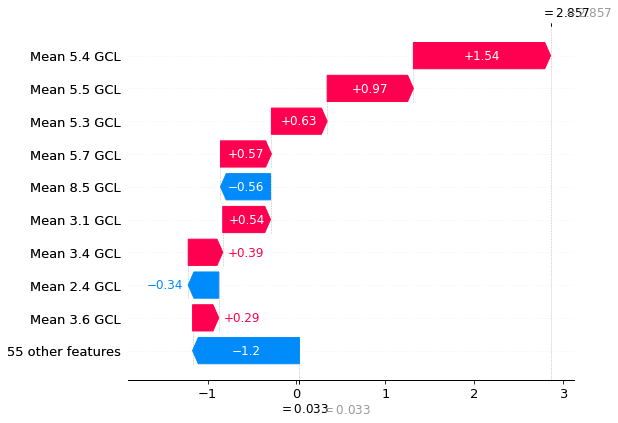} \\
\includegraphics[width=0.15\textwidth]{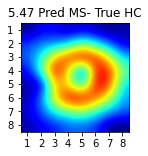} &
\includegraphics[width=0.3\textwidth]{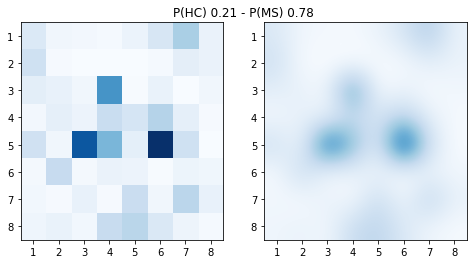} &
\includegraphics[width=0.3\textwidth]{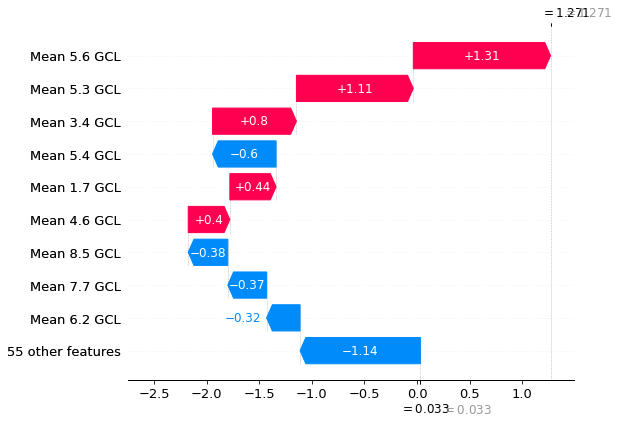} \\
\includegraphics[width=0.15\textwidth]{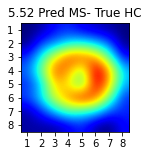} &
\includegraphics[width=0.3\textwidth]{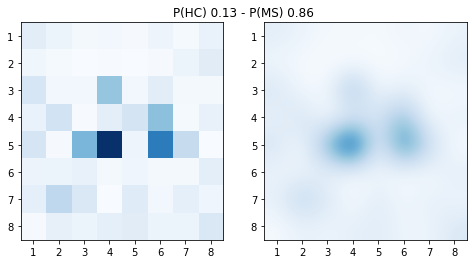} &
\includegraphics[width=0.3\textwidth]{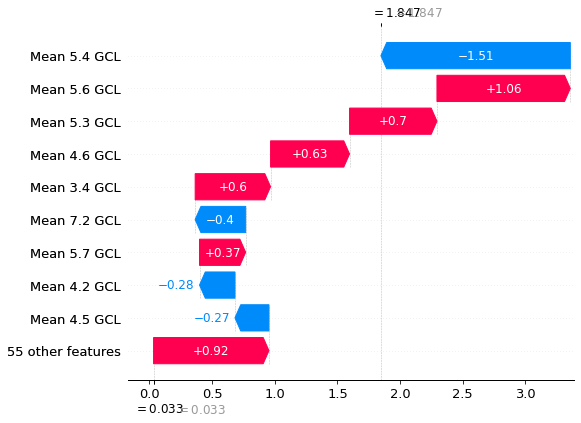} \\
\end{tabular}
\caption{Local SHAP results. Gradient boosting (XGB) best-performing model. Left, Posterior Pole (PPole) grid sample.
Middle, grid of the local SHAP values. Right, waterfall plot for assessing feature contribution towards or against multiple sclerosis (MS).}
\label{fig:SHAPXGBPPoleBestLocalFails}
\end{figure}

\begin{figure}[!htpb]
\centering
\scriptsize
RF \\
\begin{tabular}{ccc}
\includegraphics[width=0.15\textwidth]{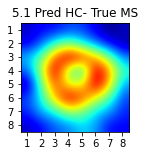} &
\includegraphics[width=0.3\textwidth]{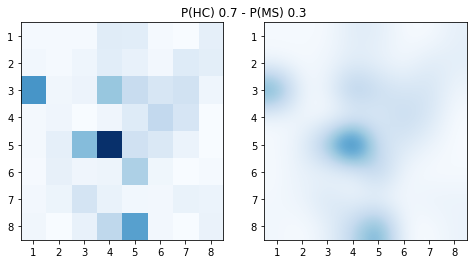} &
\includegraphics[width=0.3\textwidth]{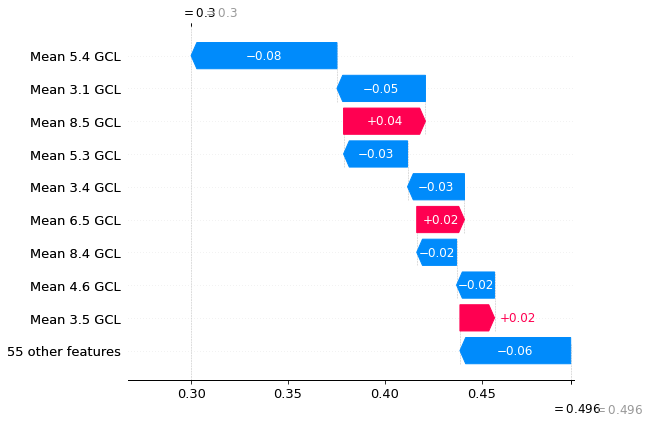} \\
\includegraphics[width=0.15\textwidth]{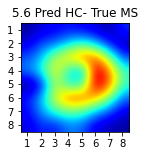} &
\includegraphics[width=0.3\textwidth]{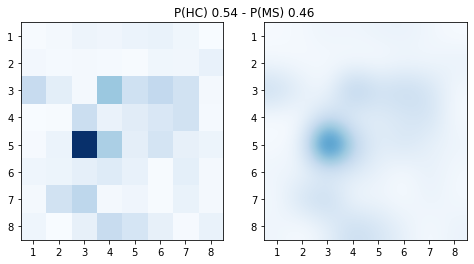} &
\includegraphics[width=0.3\textwidth]{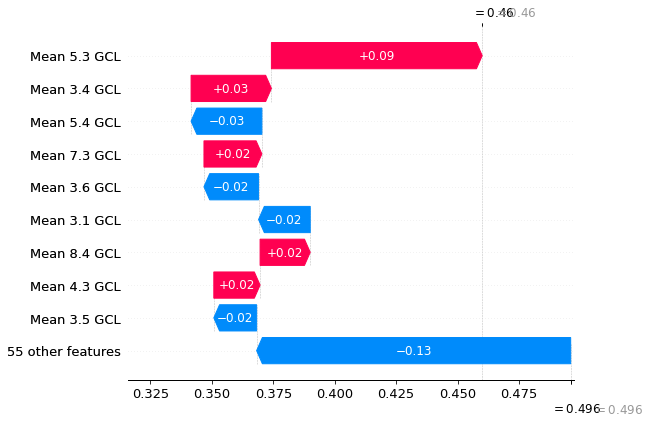} \\
\includegraphics[width=0.15\textwidth]{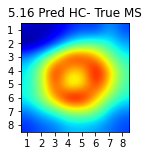} &
\includegraphics[width=0.3\textwidth]{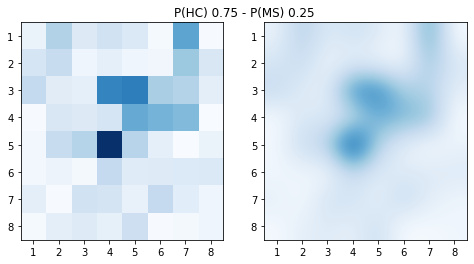} &
\includegraphics[width=0.3\textwidth]{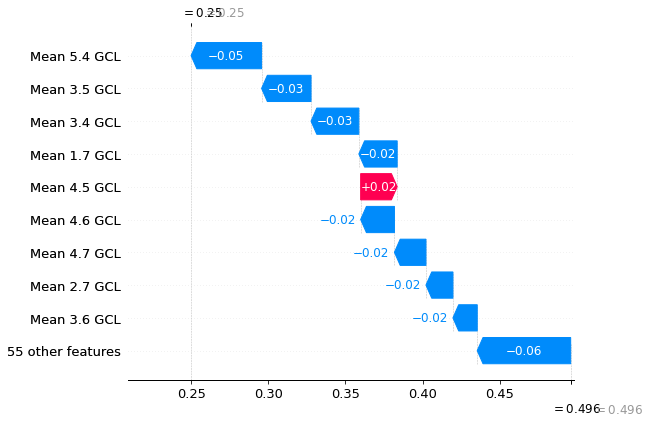} \\
\includegraphics[width=0.15\textwidth]{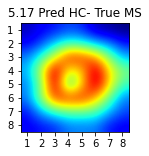} &
\includegraphics[width=0.3\textwidth]{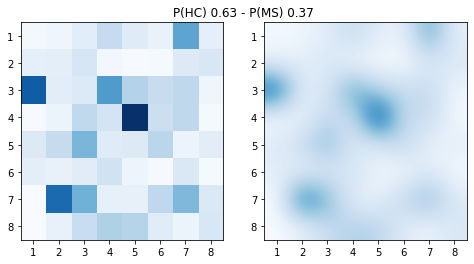} &
\includegraphics[width=0.3\textwidth]{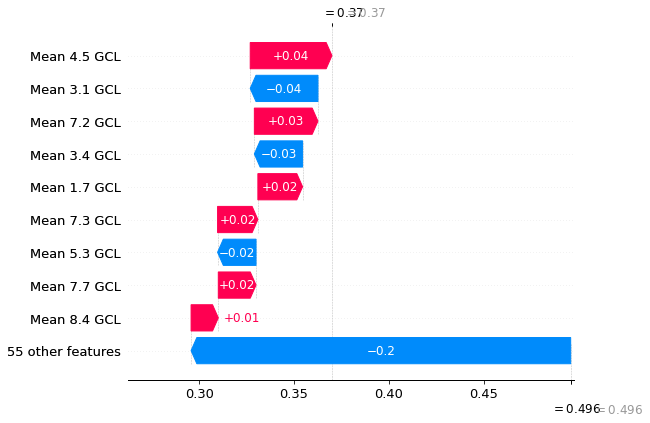} \\
\includegraphics[width=0.15\textwidth]{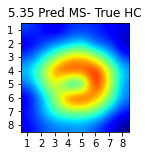} &
\includegraphics[width=0.3\textwidth]{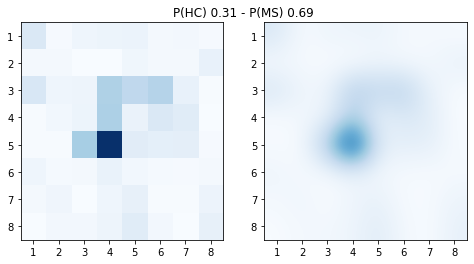} &
\includegraphics[width=0.3\textwidth]{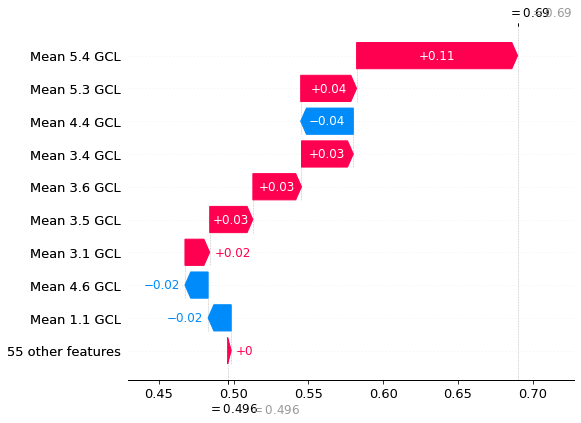} \\
\includegraphics[width=0.15\textwidth]{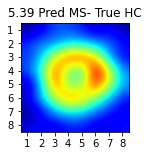} &
\includegraphics[width=0.3\textwidth]{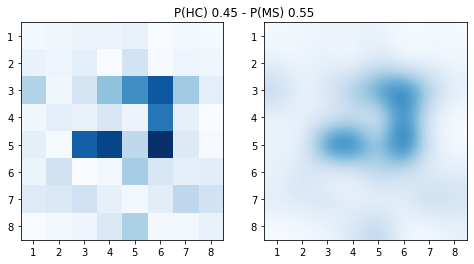} &
\includegraphics[width=0.3\textwidth]{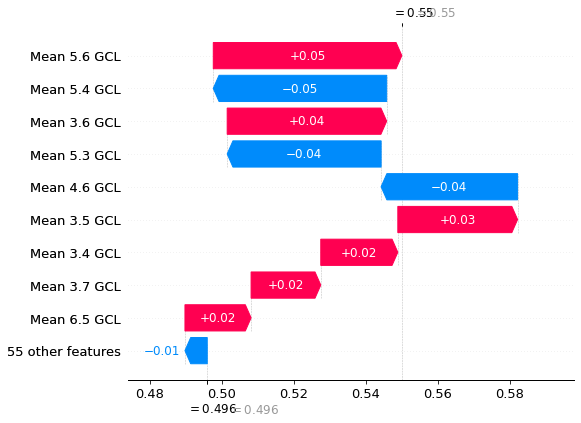} \\
\includegraphics[width=0.15\textwidth]{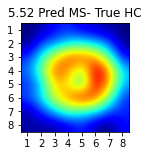} &
\includegraphics[width=0.3\textwidth]{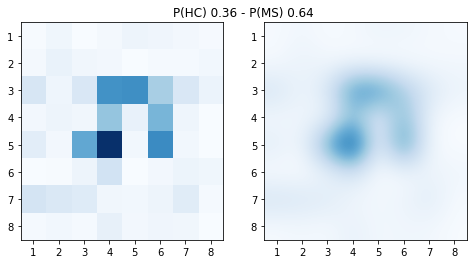} &
\includegraphics[width=0.3\textwidth]{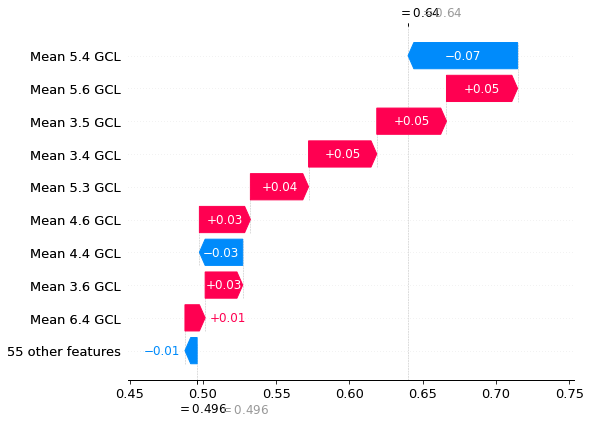} \\
\end{tabular}
\caption{Local SHAP results. Random forests (RF) best-performing model. Left, Posterior Pole (PPole) grid sample.
Middle, grid of the local SHAP values. Right, waterfall plot for assessing feature contribution towards or against multiple sclerosis (MS).
}
\label{fig:SHAPRFPPoleBestLocalFails}
\end{figure}

Figures~\ref{fig:SHAPXGBPPoleBestLocalFails} and~\ref{fig:SHAPRFPPoleBestLocalFails} show the local explainability associated
with the failed test set subjects for the best-performing XGB and RF models and PPole grid feature set.
The Supplementary Material shows the local explainability for the worst-performing XGB and RF models.
For each sample we show the PPole grid data, the grid of the corresponding SHAP values, and the waterfall plot, which indicates
the contribution of each feature toward (red) or against (blue) MS.
The corresponding confusion matrices can be found in Figure~\ref{fig:ConfusionMatrices}.

Analyzing the MS subjects that were incorrectly classified as HC for XGB, we can see that three of the MS subjects
show a shape typically found in the HC set (test set ids 5.1, 5.16, and 5.17).
Surprisingly, the 5.13 subject shows the typical break observed in the MS set, but it was classified as HC.
From the corresponding waterfall plot, we can see that feature 5.3 contributed to the probability of MS, but it was
neutralized by feature 5.4.
The same happened with the following meaningful features.
In the end, less meaningful features against the correct classification contributed to the final outcome.
The probability for HC is low, therefore the decision made by the model should not be trusted.

From the HC subjects incorrectly classified as MS by XGB, we can see that subjects 5.28 and 5.35 show the typical torus
shape break or a thinning in feature 5.4.
Subjects 5.47 and 5.52 show a similar SHAP grid, and it seems that features 5.3, 5.4 and 5.6 were mostly responsible of
the erroneous classification.
Although the overall shape is typically observed in the HC group, considering the thickness values individually
may explain the error in classification.

Similarly analyzing the subjects that were incorrectly classified as HC for the best-performing RF model, we find again subjects
5.1, 5.16, and 5.17 that showed a shape typically found in the HC group.
Subject 5.6 shows the typical torus shape break and feature 5.3. contributed to the probability of MS.
However, the remaining 55 features surprisingly summed up against the correct classification.
This may be an indication that the decision should not be trusted.

From the HC subjects that were incorrectly classified as MS, subjects 5.35 and 5.52 match with the XGB subjects.
Subject 5.39 shows a torus thinning in feature 5.6, which was considered among the most important features.
The waterfall plot shows oscillating contributions, which explains why it ended up in the incorrect outcome.
Our clinical experts indicated that this torus configuration is typical from early glaucoma.
The probabilities estimated by the model are low in all cases, therefore the decisions made by the model should not be trusted.


\subsubsection{Global EBM-based explainability}

\begin{figure}[!tpb]
\centering
\scriptsize
EBM \\
\vspace{0.25cm}
\begin{tabular}{c}
{\tiny GCL, acc 88.33} \\
\vspace{-0.5cm}
\includegraphics[width=0.5\textwidth]{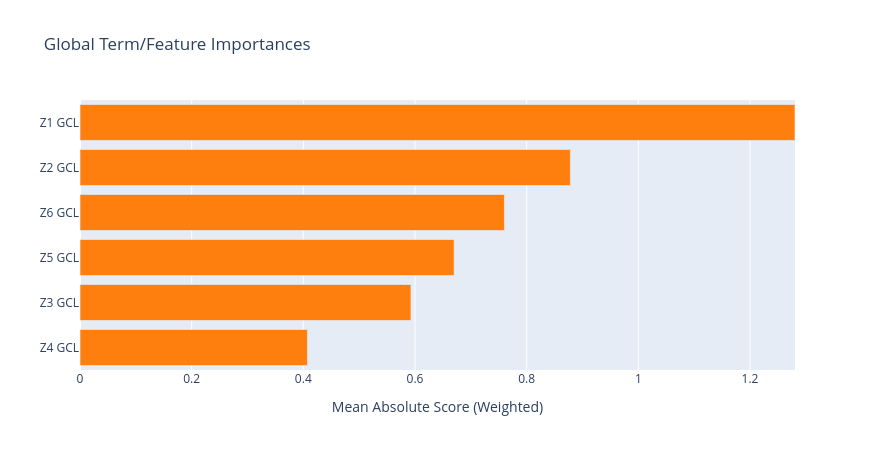} \\
\end{tabular}
\begin{tabular}{ccc}
\includegraphics[width=0.3\textwidth]{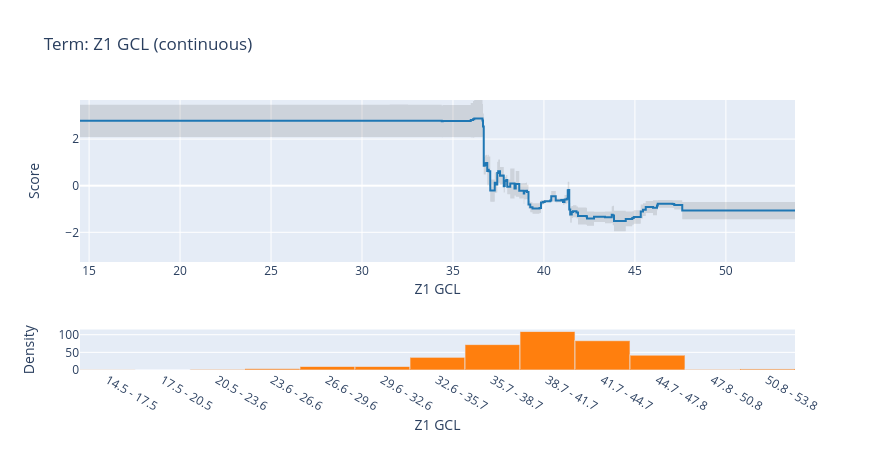} &
\includegraphics[width=0.3\textwidth]{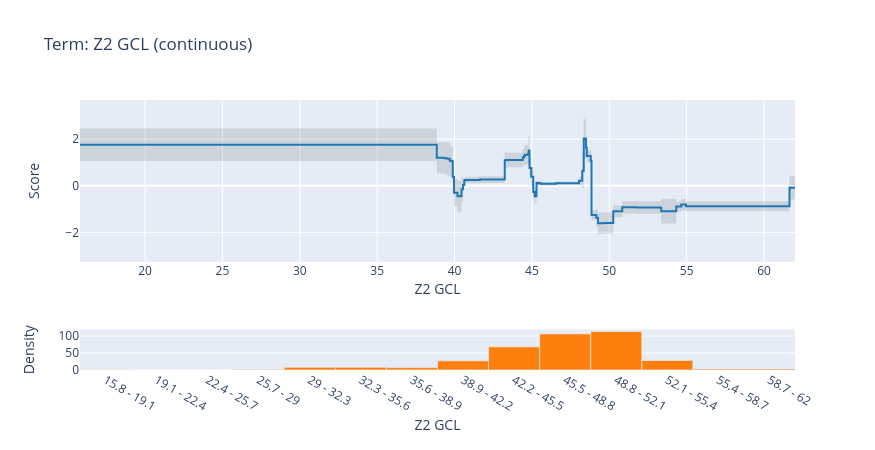} &
\includegraphics[width=0.3\textwidth]{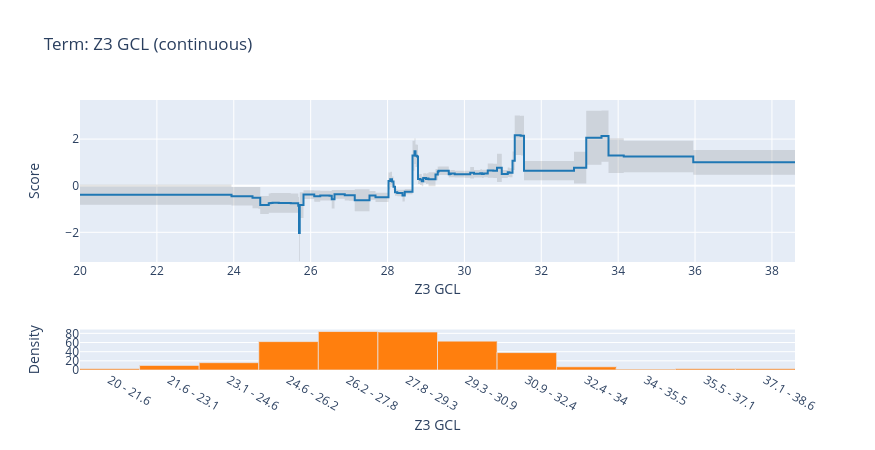} \\
\includegraphics[width=0.3\textwidth]{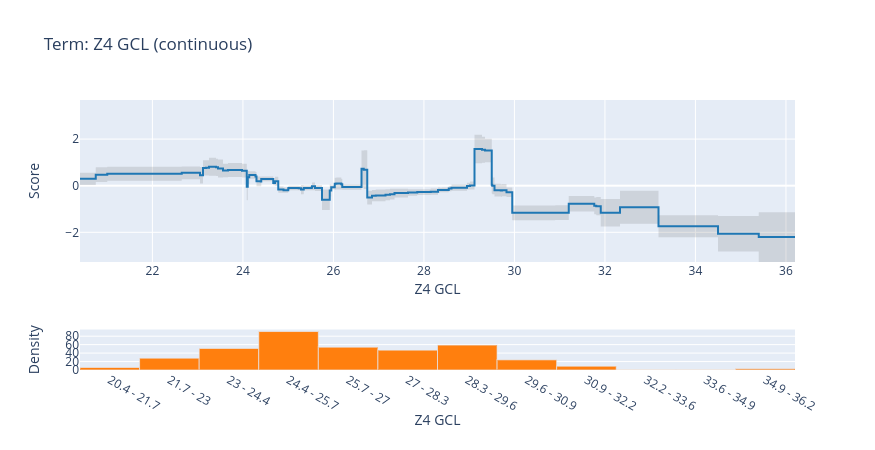} &
\includegraphics[width=0.3\textwidth]{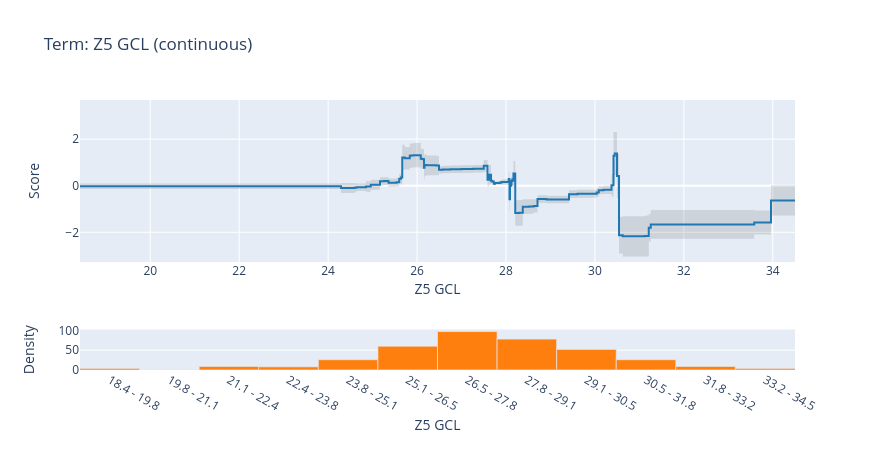} &
\includegraphics[width=0.3\textwidth]{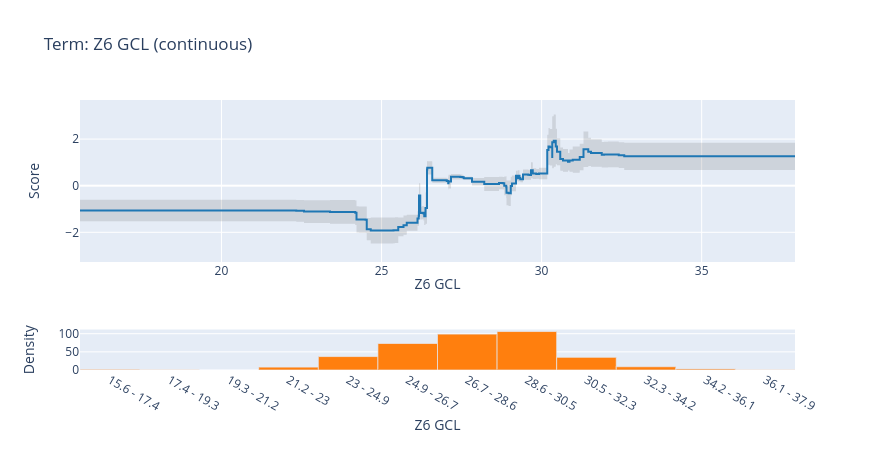} \\
\end{tabular}
\caption{Global explainable boosting machine (EBM) results. Scores for the best EBM model obtained with the Zones feature set.
Up, horizontal bar plots of the scores.
Down, curves of the score values depending on each feature value.}
\label{fig:EBMZonesBest}
\end{figure}

Figure~\ref{fig:EBMZonesBest} shows the average of the scores for the best
performing EBM model with the Zones feature set.
These values are represented as horizontal bar plots ranked by order of importance. 
In addition, we show for each feature the curves of the score values depending on the feature values.
The information driven from these curves would be analogous to the information given by SHAP with
the violin plot representation.

As with XGB and RF, zones Z1 and Z2 are the most relevant for the discrimination of MS from HC.
The order of importance matches the order of XGB.
The curves show that high values of Z1 contribute to a lower probability of MS.
The curves from Z2 show oscillating contributions from low to medium values, while the highest values
contribute to a lower probability of MS.
The pattern is similar for Z3 and Z4 while it is reversed for Z5 and Z6.
These results are analogous to the violin plots shown in Figure~\ref{fig:SHAPZonesBest}.

\begin{figure}[!htpb]
\centering
\scriptsize
EBM \\
\begin{tabular}{cc}
\includegraphics[width=0.5\textwidth]{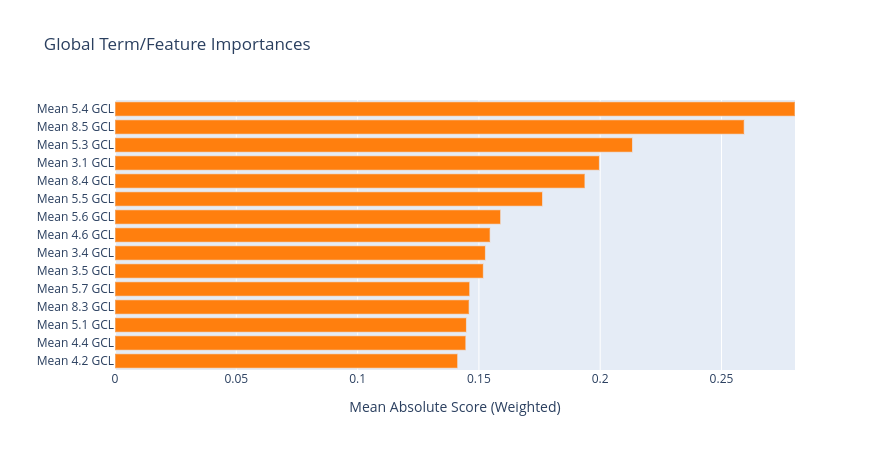} &
\includegraphics[width=0.5\textwidth]{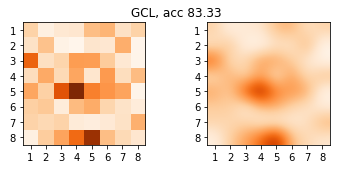} \\
\includegraphics[width=0.5\textwidth]{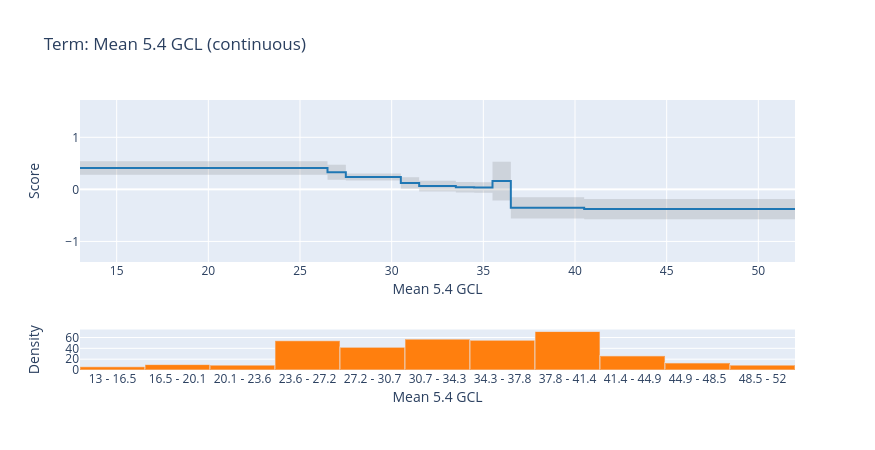} &
\includegraphics[width=0.5\textwidth]{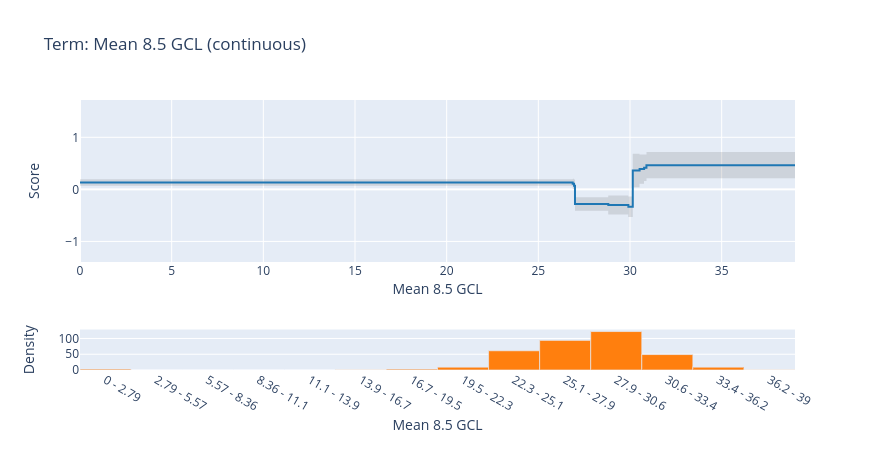} \\
\includegraphics[width=0.5\textwidth]{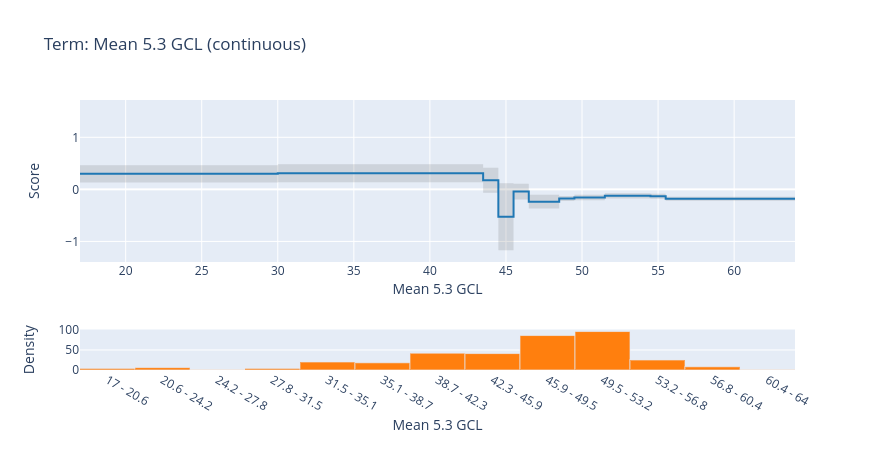} &
\includegraphics[width=0.5\textwidth]{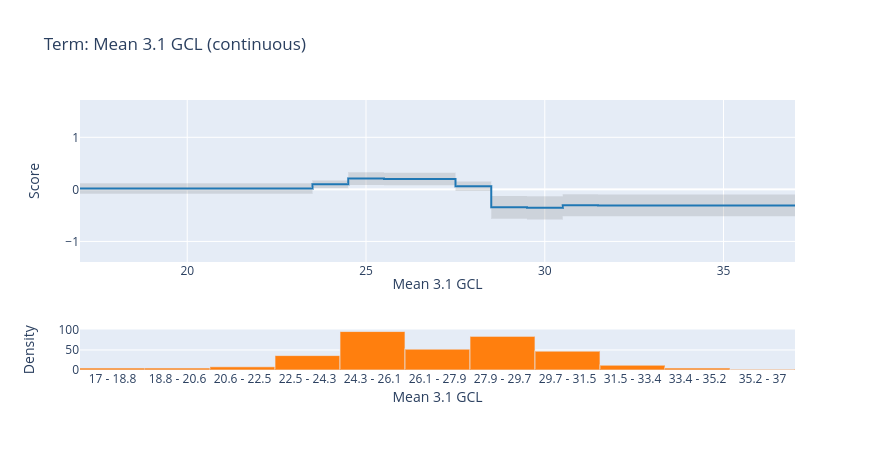} \\
\end{tabular}
\caption{Global explainable boosting machine (EBM) results. Scores for the best EBM model obtained with the posterior pole (PPole) grid feature set.
Up, horizontal bar plots of the scores and grids.
Down, curves of the score values of the top-four feature set.}
\label{fig:EBMPPPoleBest}
\end{figure}


Figure~\ref{fig:EBMPPPoleBest} shows the scores for the best-performing EBM model with the PPole grid feature set.
Additionally to the bar plots, we show the score values on the $8 \times 8$ grid.
As with XGB and RF, features 5.4 and 5.3 are in the top-three of relevance for the discrimination of MS from HC.
It drives our attention that features in the boundary of the grid such as 8.5, 3.1, and 8.4 occupy the top positions.
The grid shows that feature importance is distributed among features 5.4 and 5.3 and the torus-like shape.
However, feature importance is also located in two points of the boundary.
The curves for features 5.4 and 5.3 show that high values contribute to a low probability of MS.
The same happens with the curves for feature 3.1.
On the contrary, the curves for feature 8.5 show that high values contribute to a high probability of MS.
We find again a possible bias of the model that considers this location useful for the discrimination between groups.

\subsubsection{Local EBM-based explainability}

\begin{figure}[!htpb]
\centering
\scriptsize
EBM \\
\begin{tabular}{ccc}
\includegraphics[width=0.15\textwidth]{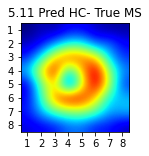} &
\includegraphics[width=0.3\textwidth]{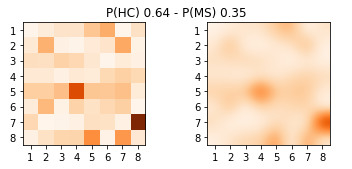} &
\includegraphics[width=0.25\textwidth]{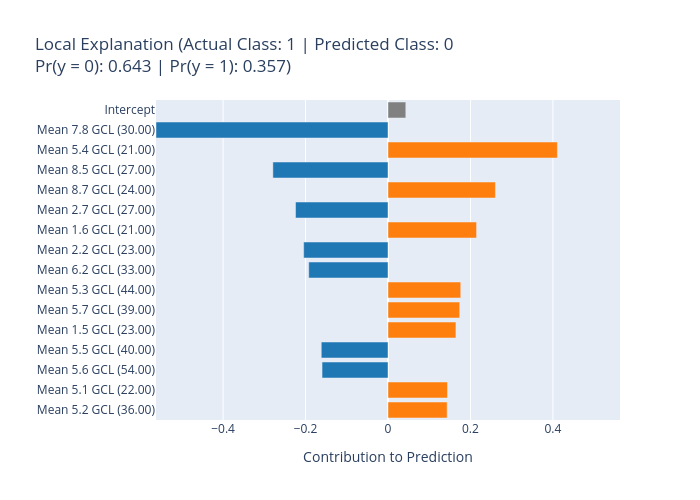} \\
\includegraphics[width=0.15\textwidth]{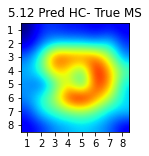} &
\includegraphics[width=0.3\textwidth]{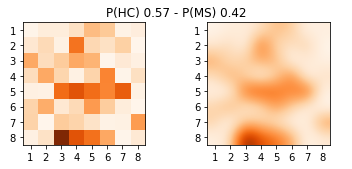} &
\includegraphics[width=0.25\textwidth]{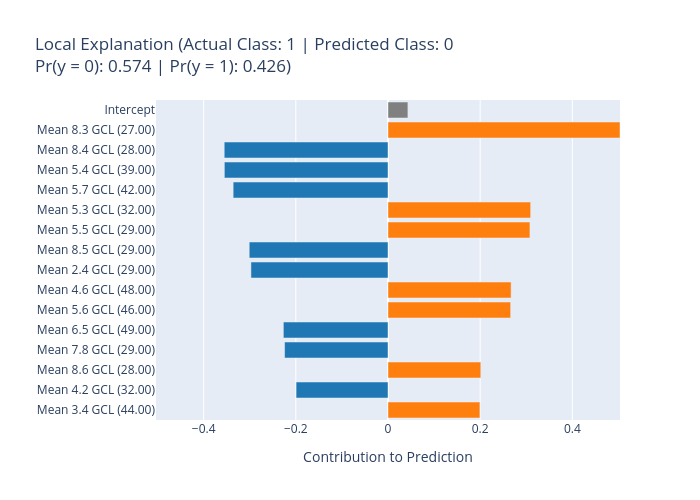} \\
\includegraphics[width=0.15\textwidth]{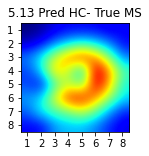} &
\includegraphics[width=0.3\textwidth]{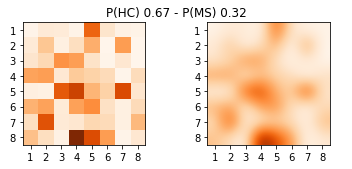} &
\includegraphics[width=0.25\textwidth]{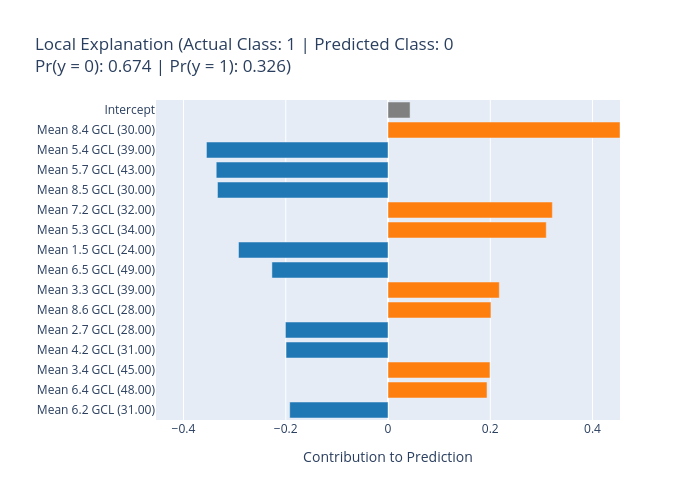} \\
\includegraphics[width=0.15\textwidth]{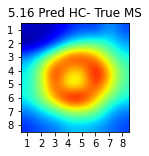} &
\includegraphics[width=0.3\textwidth]{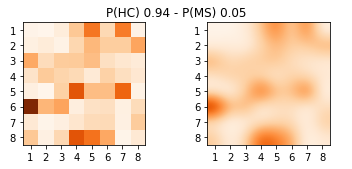} &
\includegraphics[width=0.25\textwidth]{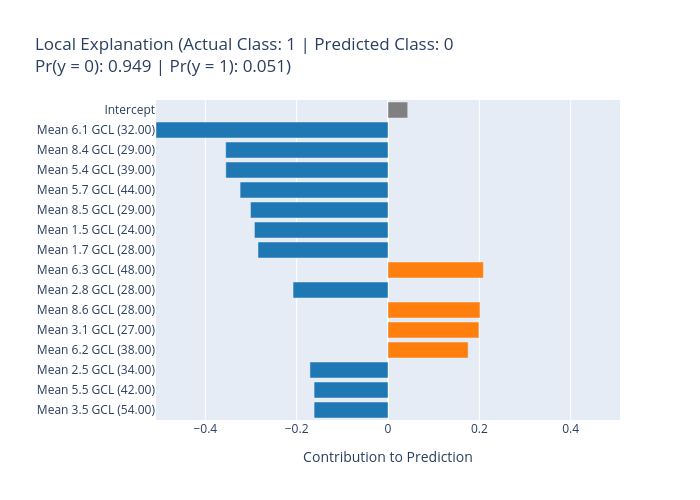} \\
\includegraphics[width=0.15\textwidth]{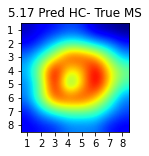} &
\includegraphics[width=0.3\textwidth]{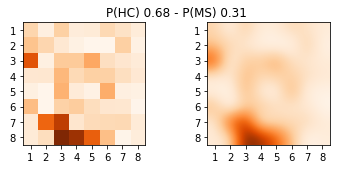} &
\includegraphics[width=0.25\textwidth]{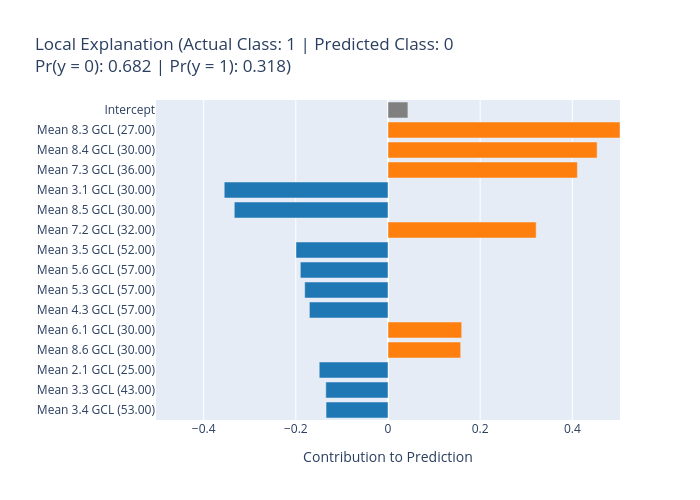} \\
\includegraphics[width=0.15\textwidth]{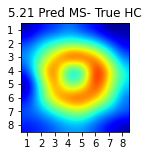} &
\includegraphics[width=0.3\textwidth]{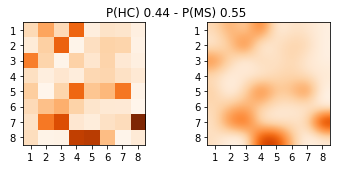} &
\includegraphics[width=0.25\textwidth]{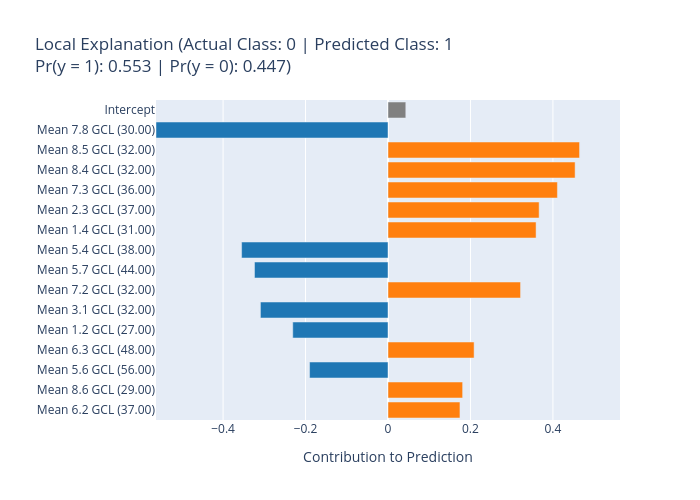} \\
\includegraphics[width=0.15\textwidth]{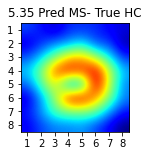} &
\includegraphics[width=0.3\textwidth]{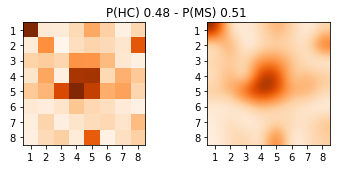} &
\includegraphics[width=0.25\textwidth]{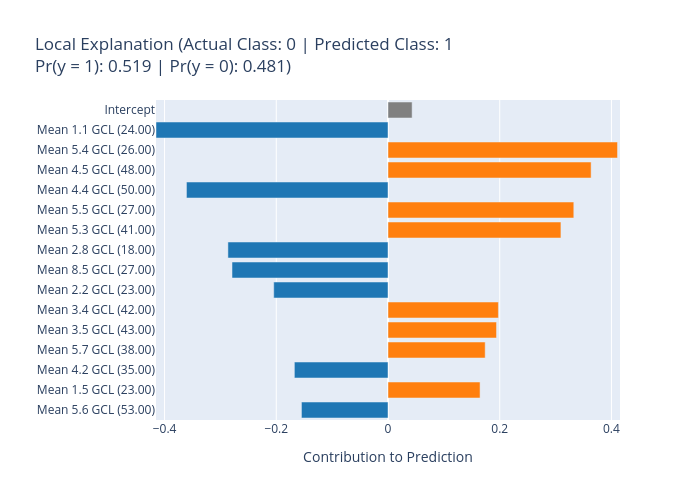} \\
\includegraphics[width=0.15\textwidth]{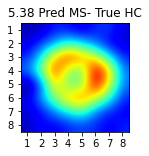} &
\includegraphics[width=0.3\textwidth]{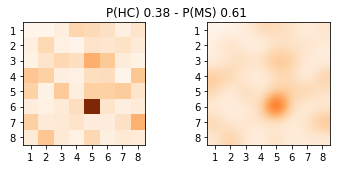} &
\includegraphics[width=0.25\textwidth]{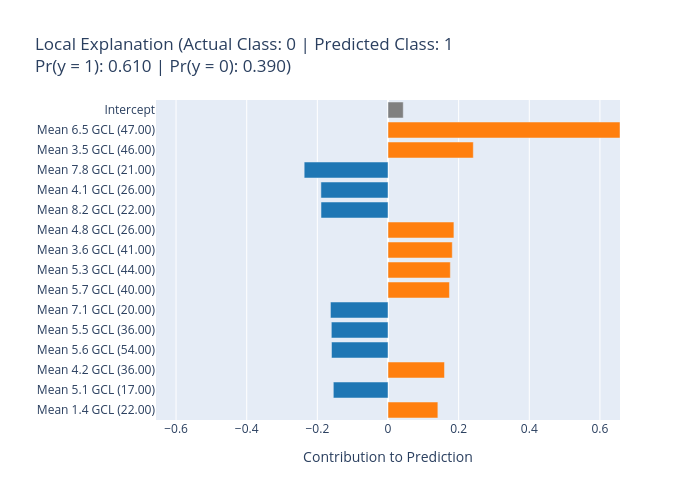} \\
\includegraphics[width=0.15\textwidth]{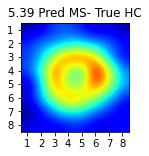} &
\includegraphics[width=0.3\textwidth]{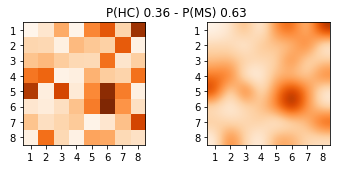} &
\includegraphics[width=0.25\textwidth]{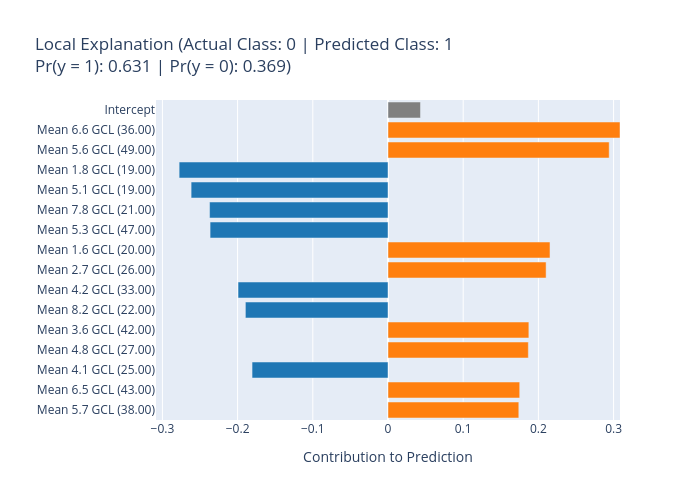} \\
\includegraphics[width=0.15\textwidth]{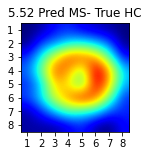} &
\includegraphics[width=0.3\textwidth]{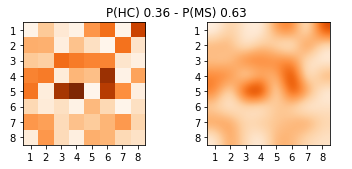} &
\includegraphics[width=0.25\textwidth]{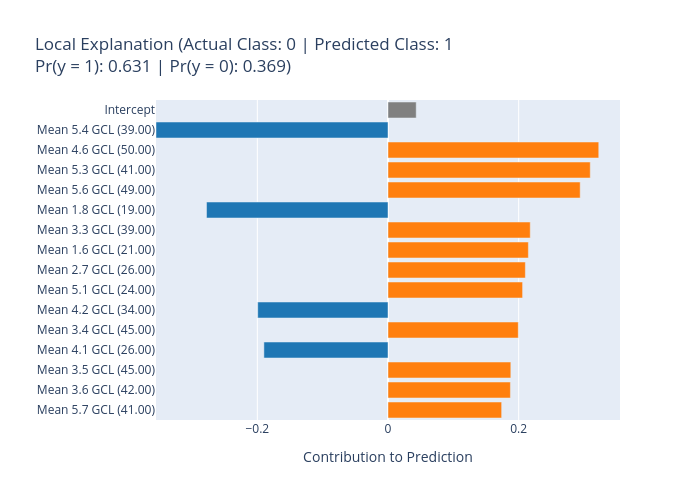} \\
\end{tabular}
\caption{Local explainable boosting machine (EBM) results. EBM best-performing model. For each subject, left figure shows the posterior pole (PPole) grid sample,
middle figure shows the grids, and right figure shows the feature contribution towards or against multiple sclerosis (MS).}
\label{fig:EBMPPoleBestLocalFails}
\end{figure}

Figure~\ref{fig:EBMPPoleBestLocalFails} shows the local explainability associated with the failed test set subjects
for the best-performing EBM model and PPole grid feature set.
For each sample, we show the PPole grid data, the 2D grid of scores, and the local importance scores.
The corresponding confusion matrix can be found in Figure~\ref{fig:ConfusionMatrices}.
The Supplementary Material shows the local explainability for the worst-performing EBM model.

Analyzing the MS subjects incorrectly classified as HC for EBM, we find subjects 5.16 and 5.17 that showed a shape typically
found in the HC group.
Subjects 5.11, 5.12, and 5.13 should be clearly identified as MS patients, but the importance given by the model to the boundary
points 7.8, 8.3, and 8.4 contributed to an incorrect decision.
For the HC subjects incorrectly classified as MS, subject 5.21 has been incorrectly classified using information from feature
points 7.8, 8.5, and 8.4 and subject 5.35 from point 1.1.
The remaining subjects show patterns observed in the MS population, and the model attributed the greatest importance to
the locations where these patterns are found, e.g., subject 5.38 and features 6.5 and 3.5; subject 5.39 and features 6.6. and
5.6; subject 5.52 and features 5.4, 4.6, and 5.3.

\subsubsection{EBM with interactions}

\begin{figure}[!htpb]
\centering
\scriptsize
EBM + i \\
\begin{tabular}{cc}
{\tiny GCL, acc 90.32} & {\tiny GCL, acc 83.87} \\
\includegraphics[width=0.5\textwidth]{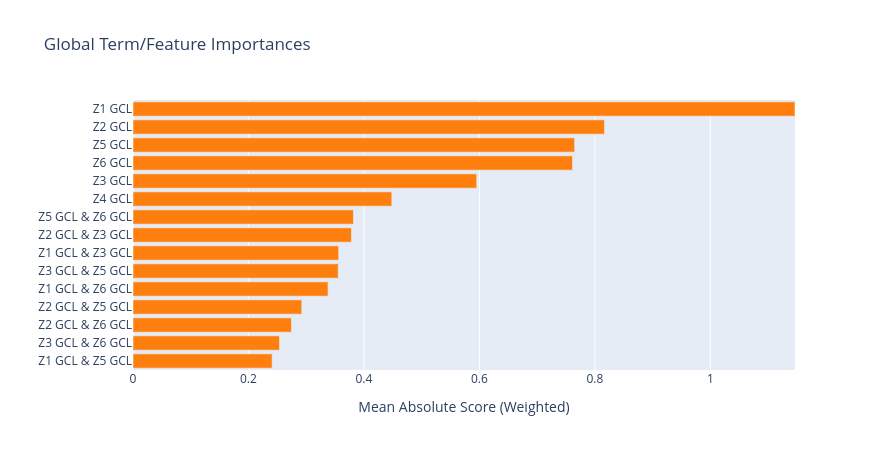} &
\includegraphics[width=0.5\textwidth]{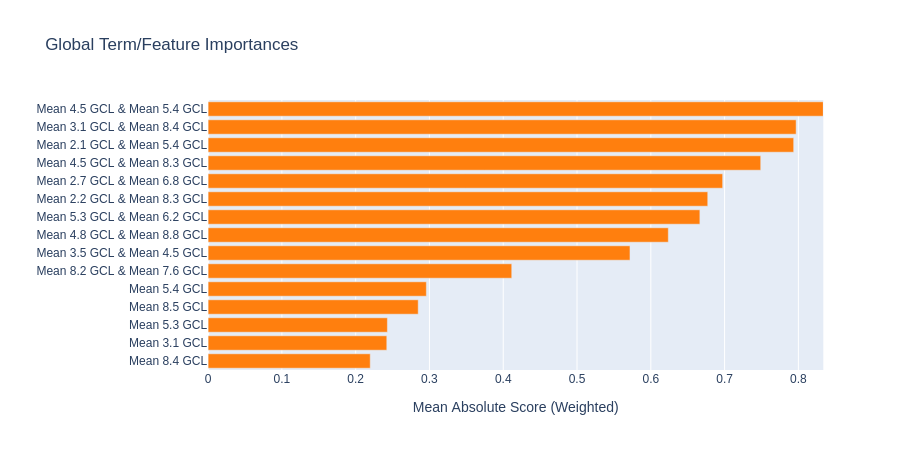} \\
\includegraphics[width=0.5\textwidth]{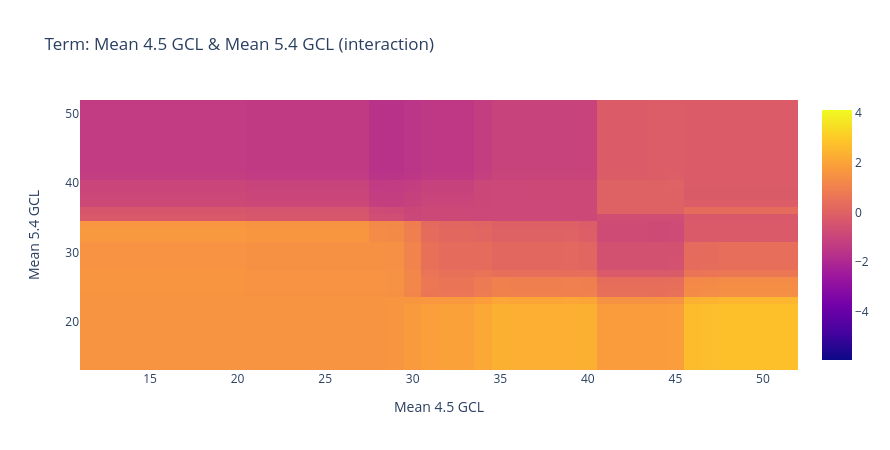} &
\includegraphics[width=0.5\textwidth]{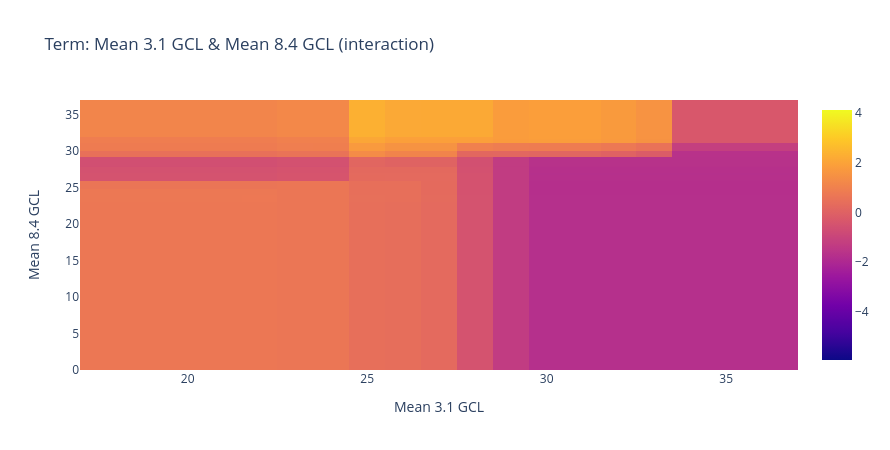} \\
\includegraphics[width=0.5\textwidth]{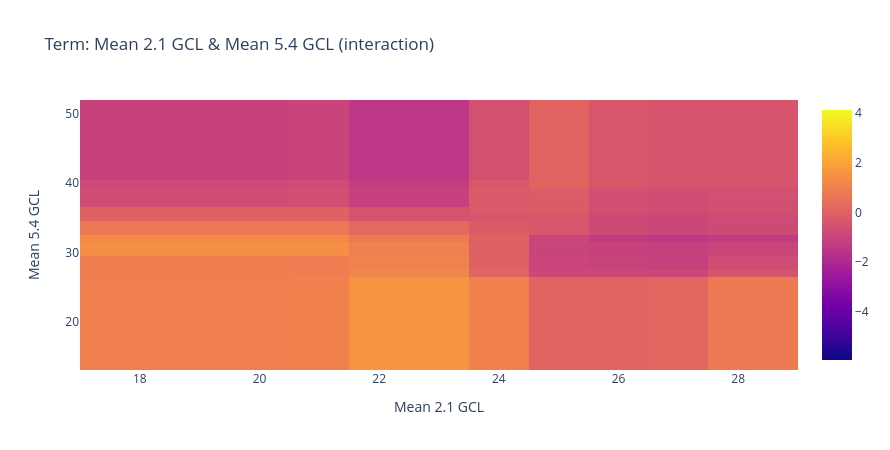} &
\includegraphics[width=0.5\textwidth]{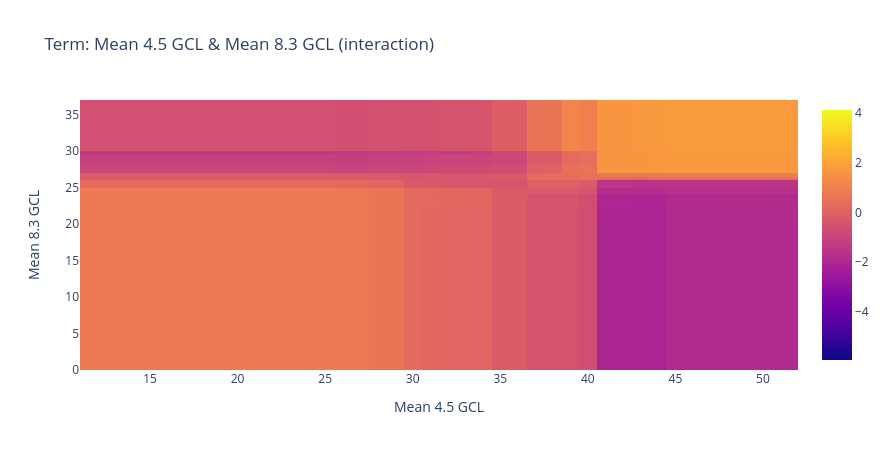} \\
\end{tabular}
\caption{Global explainable boosting machine with interactions (EBM + i) results. Up, horizontal bars of the scores for the best EBM + i model obtained with
Zones and posterior pole (PPole) grid feature sets.
Middle and down, interaction maps of the top-four feature set for the PPole grid feature set.
}
\label{fig:EBMiPPPoleBest}
\end{figure}

Figure~\ref{fig:EBMiPPPoleBest} shows the average of the scores for the best performing EBM + i model with the Zones
and PPole grid feature sets.
While for the Zones feature set the importance rank is similar between EBM and EBM + i models, pairwise interactions
occupy the top-ten positions for the EBM + i model with the PPole grid feature set.
Figure~\ref{fig:EBMiPPPoleBest} also shows the interaction plot for the top-four positions.
The top-one interaction is between points 4.5 and 5.4. This is consistent with the explanations given by XGB and RF.
The interaction plot shows that for low values of thickness in 4.5 and 5.4 the probability of MS is high, as expected.
For low values in 5.4 and high values in 4.5, the probability of MS is high.
This means that the model uses the torus shape break or thinning in this location for taking the decision.
On the contrary, for high values in 5.4 and low values in 4.5, then the probability of MS is low.
Finally, for high values in 5.4 and 4.5 the probability of MS is nearly zero.
In this case, these features are not relevant for the model to distinguish between groups.

The second interaction is between the boundary points 3.1 and 8.4.
As happened in the case of the top-one interaction, for low values of both features, the probability of MS is high,
and for high values of both features the probability of MS is nearly zero.
For low values in 8.4 and high values of 3.1 the probability of MS is low.
For high values in 8.4 and low values in 3.1 the probability of MS is high.
Recalling the suspected bias found in points 8.4 and 8.5 for the model without interactions,
it seems that in this case the bias is shown in this interaction of highly unlikely related features.
The same reasoning can be applied to the interaction between points 2.1 and 5.4 and between points 4.5 and
8.3.

\begin{figure}[!htpb]
\centering
\scriptsize
EBM + i \\
\begin{tabular}{cccc}
\includegraphics[width=0.15\textwidth]{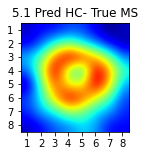} &
\includegraphics[width=0.3\textwidth]{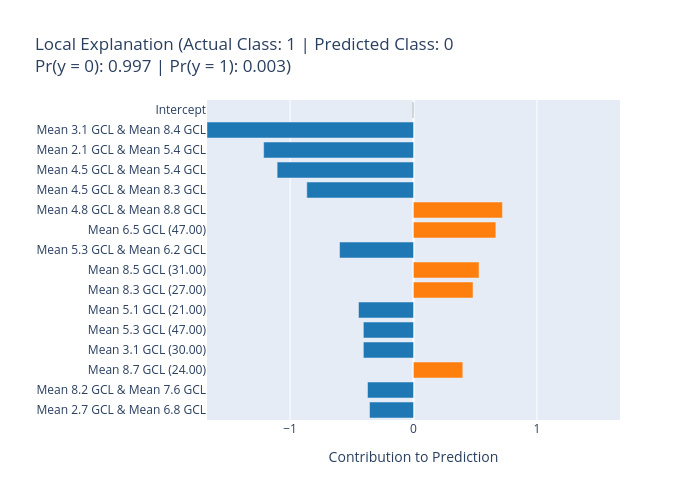} &
\includegraphics[width=0.15\textwidth]{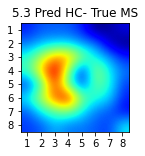} &
\includegraphics[width=0.3\textwidth]{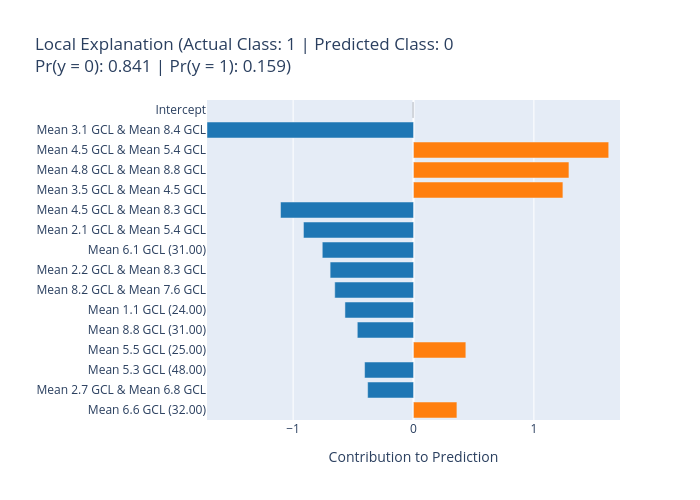} \\
\includegraphics[width=0.15\textwidth]{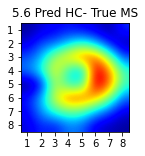} &
\includegraphics[width=0.3\textwidth]{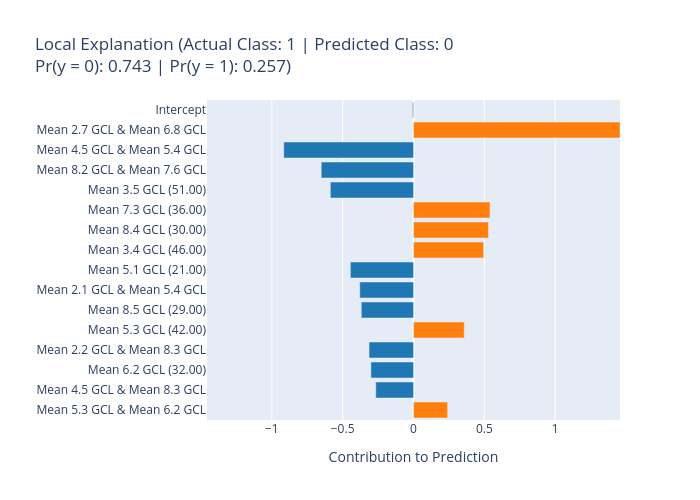} &
\includegraphics[width=0.15\textwidth]{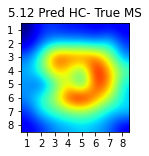} &
\includegraphics[width=0.3\textwidth]{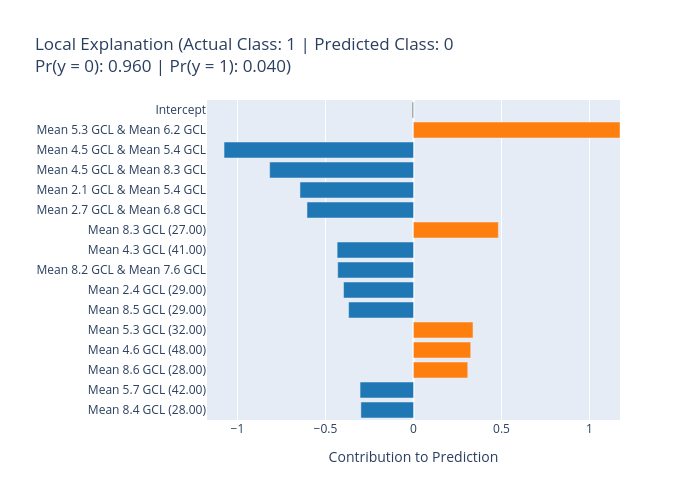} \\
\includegraphics[width=0.15\textwidth]{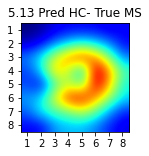} &
\includegraphics[width=0.3\textwidth]{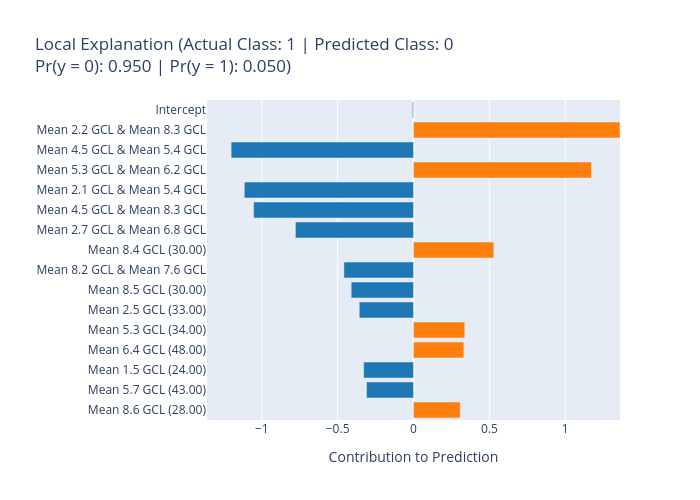} &
\includegraphics[width=0.15\textwidth]{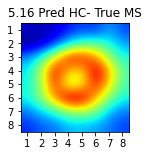} &
\includegraphics[width=0.3\textwidth]{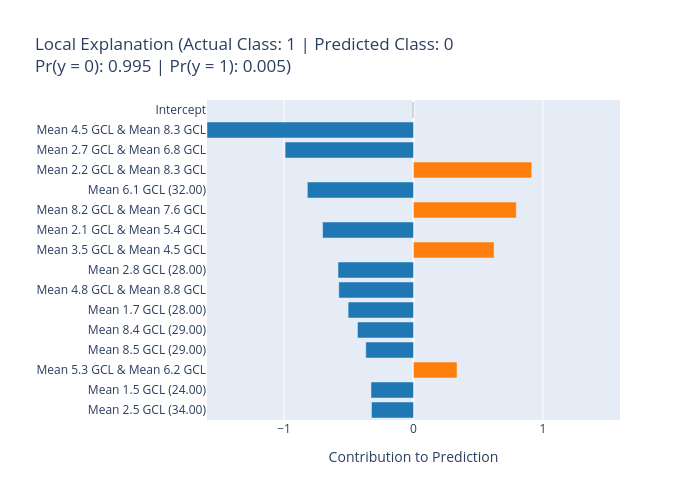} \\
\includegraphics[width=0.15\textwidth]{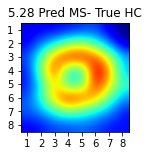} &
\includegraphics[width=0.3\textwidth]{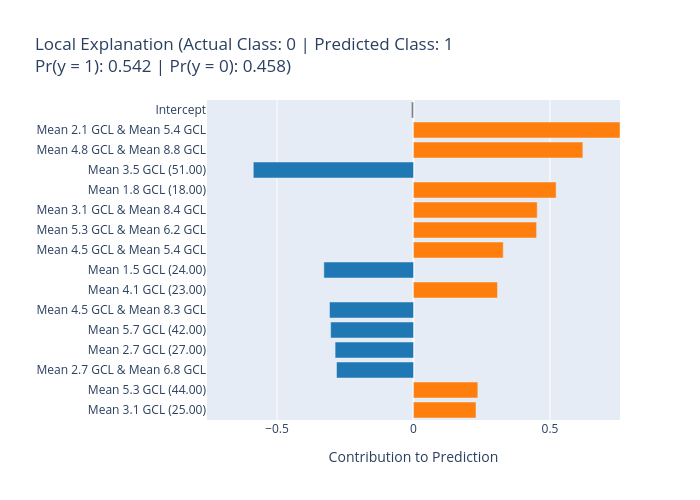} &
\includegraphics[width=0.15\textwidth]{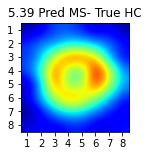} &
\includegraphics[width=0.3\textwidth]{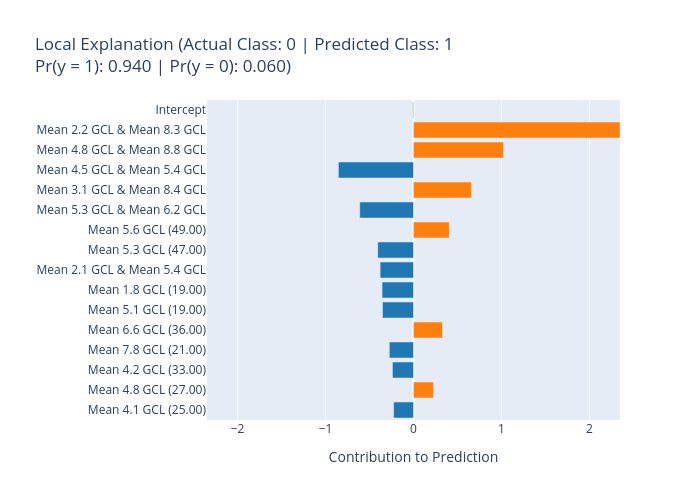} \\
\end{tabular}
\caption{Local explainable boosting machine with interactions (EBM + i) results. EBM + i best-performing model. For each subject, left figure shows the posterior pole (PPole) grid sample and right figure shows the feature contribution towards or against multiple scleroris (MS).}
\label{fig:EBMiPPoleBestLocalFails}
\end{figure}

Figure~\ref{fig:EBMiPPoleBestLocalFails} shows the local explainability associated with the failed test set subjects
for the best-performing EBM + i model and PPole grid feature set.
Both EBM and EBM + i models failed in subjects 5.12, 5.13, 5.16, and 5.39.
It seems that the most important features involving boundary points were the responsible of the failed decisions.

\subsubsection{Global explainability: layers and models}
\label{subsec:Grids}

\begin{figure}[!t]
\centering
\scriptsize
\begin{tabular}{ccc}
XGB & RF & EBM \\
\includegraphics[width=0.30\textwidth]{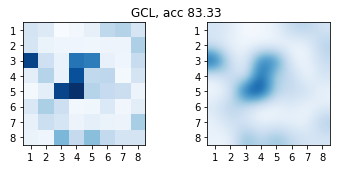} &
\includegraphics[width=0.30\textwidth]{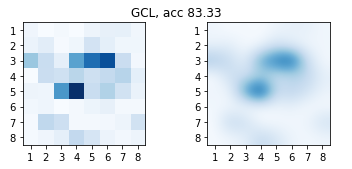} &
\includegraphics[width=0.30\textwidth]{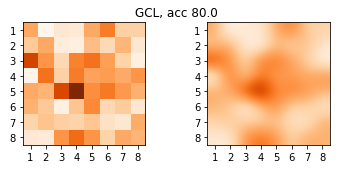}
\\
\includegraphics[width=0.30\textwidth]{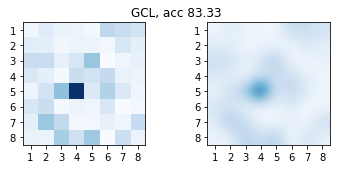} &
\includegraphics[width=0.30\textwidth]{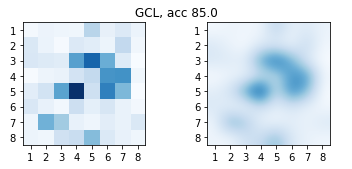} &
\includegraphics[width=0.30\textwidth]{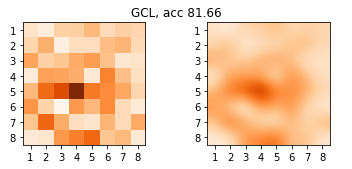}
\\
\includegraphics[width=0.30\textwidth]{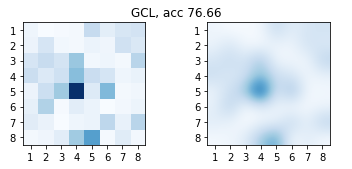} &
\includegraphics[width=0.30\textwidth]{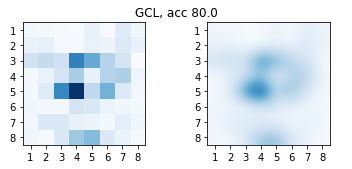} &
\includegraphics[width=0.30\textwidth]{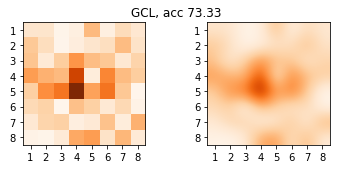}
\\
\includegraphics[width=0.30\textwidth]{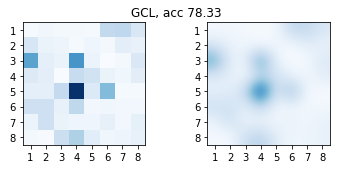} &
\includegraphics[width=0.30\textwidth]{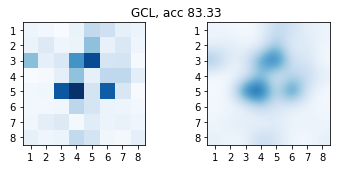} &
\includegraphics[width=0.30\textwidth]{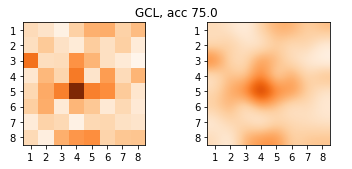}
\\
\includegraphics[width=0.30\textwidth]{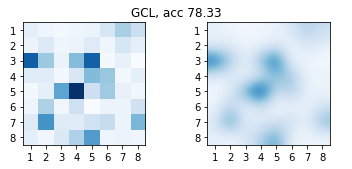} &
\includegraphics[width=0.30\textwidth]{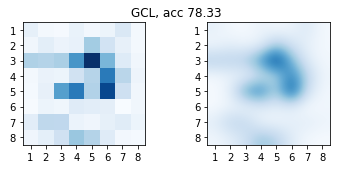} &
\includegraphics[width=0.30\textwidth]{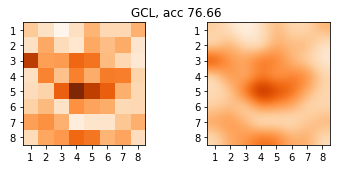}
\\
\includegraphics[width=0.30\textwidth]{./XGB/Results_CBM2023/XGB_GCL_Both_Age20-60_PPole_Universal/test_5_SHAPGrid.png} &
\includegraphics[width=0.30\textwidth]{./RF/Results_CBM2023/RF_GCL_Both_Age20-60_PPole_Universal/test_5_SHAPGrid.png} &
\includegraphics[width=0.30\textwidth]{./EBMII/Results_CBM2023/EBM_GCL_Both_Age20-60_PPole_Universal/test_5_EBMGrid.png}
\\
\includegraphics[width=0.30\textwidth]{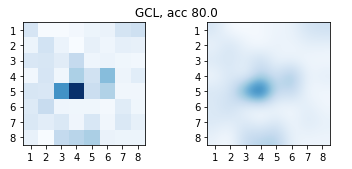} &
\includegraphics[width=0.30\textwidth]{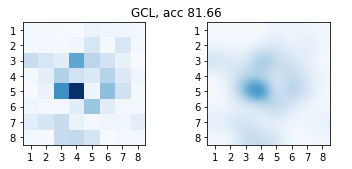} &
\includegraphics[width=0.30\textwidth]{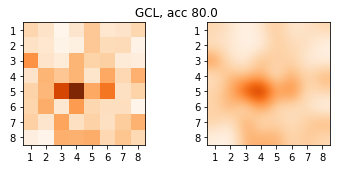}
\\
\includegraphics[width=0.30\textwidth]{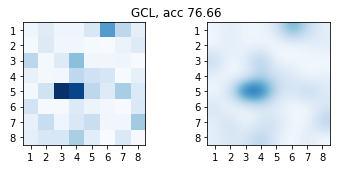} &
\includegraphics[width=0.30\textwidth]{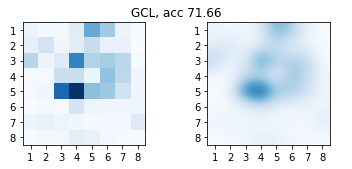} &
\includegraphics[width=0.30\textwidth]{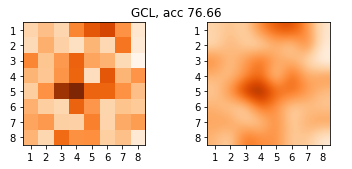}
\\
\includegraphics[width=0.30\textwidth]{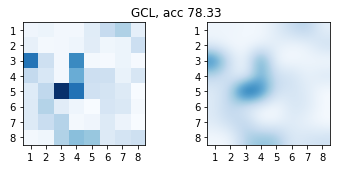} &
\includegraphics[width=0.30\textwidth]{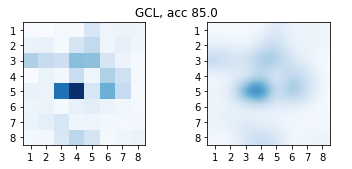} &
\includegraphics[width=0.30\textwidth]{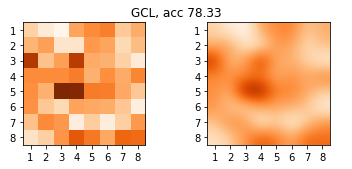}
\\
\includegraphics[width=0.30\textwidth]{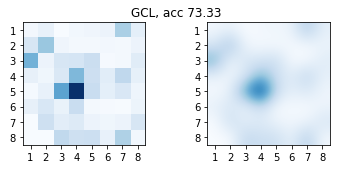} &
\includegraphics[width=0.30\textwidth]{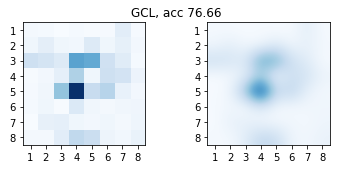} &
\includegraphics[width=0.30\textwidth]{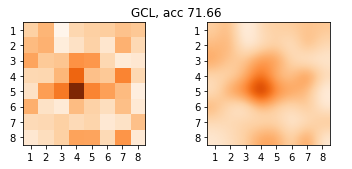}
\\
\end{tabular}
\caption{Global SHAP and explainable boosting machine (EBM) results. Ganglion cell layer (GCL). SHAP and EBM grids obtained with the models of
the ten-fold cross-validation. }
\label{fig:GCLSHAPGrids}
\end{figure}

\begin{figure}[!t]
\centering
\scriptsize
\begin{tabular}{ccc}
XGB & RF & EBM \\
\includegraphics[width=0.30\textwidth]{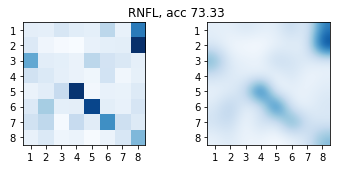} &
\includegraphics[width=0.30\textwidth]{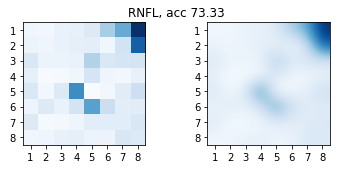} &
\includegraphics[width=0.30\textwidth]{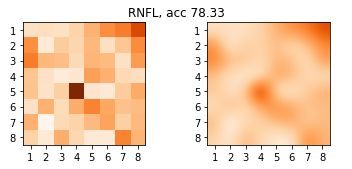}
\\
\includegraphics[width=0.30\textwidth]{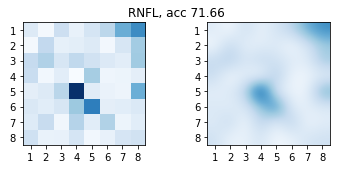} &
\includegraphics[width=0.30\textwidth]{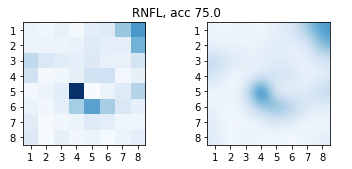} &
\includegraphics[width=0.30\textwidth]{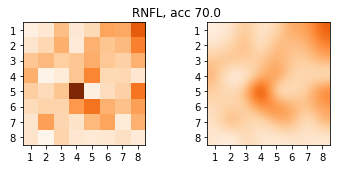}
\\
\includegraphics[width=0.30\textwidth]{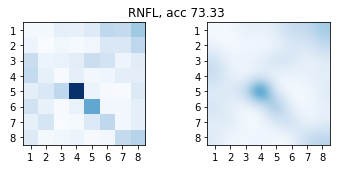} &
\includegraphics[width=0.30\textwidth]{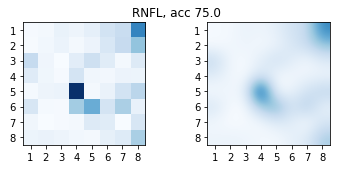} &
\includegraphics[width=0.30\textwidth]{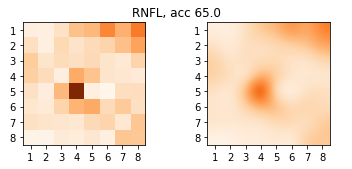}
\\
\includegraphics[width=0.30\textwidth]{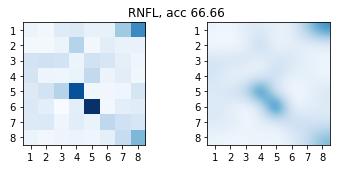} &
\includegraphics[width=0.30\textwidth]{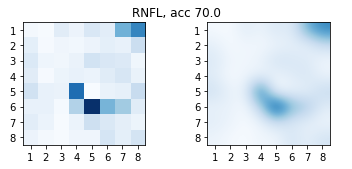} &
\includegraphics[width=0.30\textwidth]{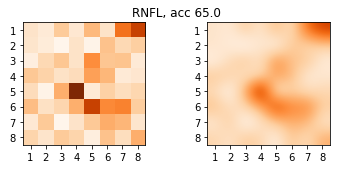}
\\
\includegraphics[width=0.30\textwidth]{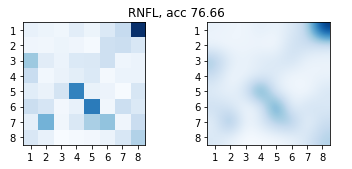} &
\includegraphics[width=0.30\textwidth]{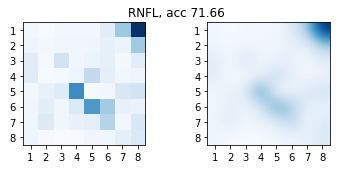} &
\includegraphics[width=0.30\textwidth]{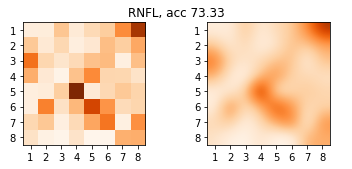}
\\
\includegraphics[width=0.30\textwidth]{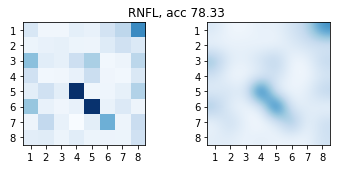} &
\includegraphics[width=0.30\textwidth]{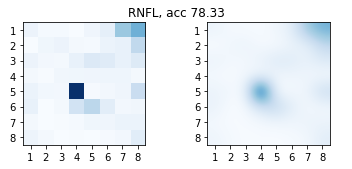} &
\includegraphics[width=0.30\textwidth]{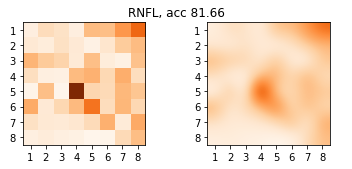}
\\
\includegraphics[width=0.30\textwidth]{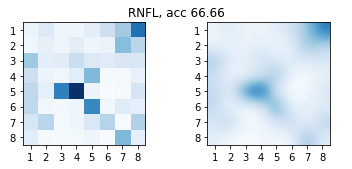} &
\includegraphics[width=0.30\textwidth]{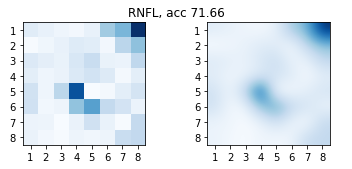} &
\includegraphics[width=0.30\textwidth]{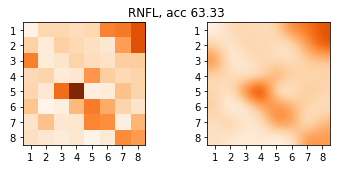}
\\
\includegraphics[width=0.30\textwidth]{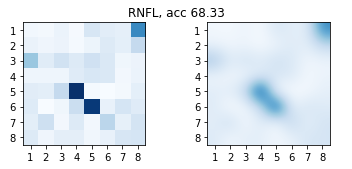} &
\includegraphics[width=0.30\textwidth]{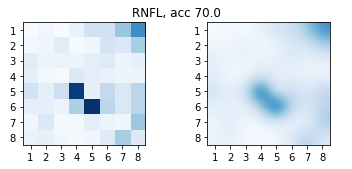} &
\includegraphics[width=0.30\textwidth]{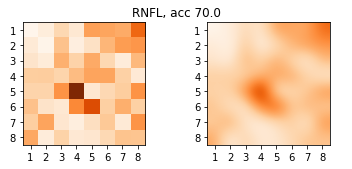}
\\
\includegraphics[width=0.30\textwidth]{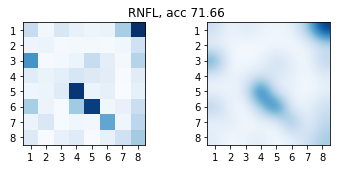} &
\includegraphics[width=0.30\textwidth]{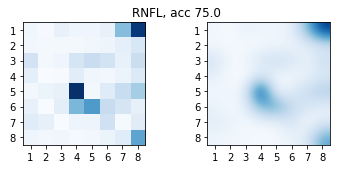} &
\includegraphics[width=0.30\textwidth]{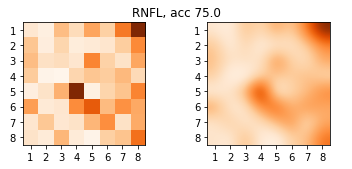}
\\
\includegraphics[width=0.30\textwidth]{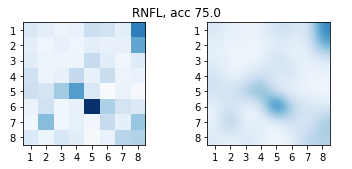} &
\includegraphics[width=0.30\textwidth]{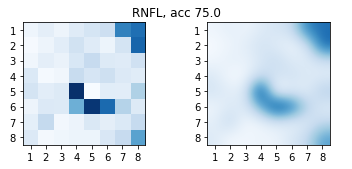} &
\includegraphics[width=0.30\textwidth]{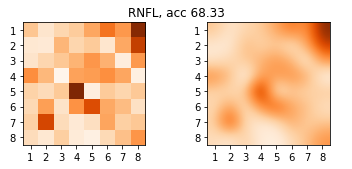}
\\
\end{tabular}
\caption{Global SHAP and explainable boosting machine (EBM) results. Retinal nerve fiber layer (RNFL). SHAP and EBM grids obtained
with the models of the ten-fold cross-validation. }
\label{fig:RNFLSHAPGrids}
\end{figure}

To finish our study, we provide a comparison of the global SHAP values and EBM scores from the PPole grid
feature set for the GCL and the RNFL layers. Figures~\ref{fig:GCLSHAPGrids} and~\ref{fig:RNFLSHAPGrids} show the grids obtained
for the models generated with the ten test sets considered in our work.
For the GCL feature set,
all the models give consistently importance to the location of the 5.4 feature.
For XGB this seems to be the single most remarkable feature in the majority of models.
RF shows importance in several locations of the torus ridge.
EBM also shows importance in several locations of the torus ridge but it is also giving importance to other
locations.
We realize that the majority of models find feature 3.1 as important, and this becomes more apparent for XGB and EBM.
EBM tends to remark as important many more points on the boundary.

For the RNFL feature set, all models consistently point out to feature 5.4.
Indeed, we can appreciate two or three diagonally aligned points consistently remarked by all the models:
feature 5.4, 6.5, and 7.6.
These points correspond to the locations of some of the nerve bundle fibers (see Figure 1 from~\cite{Qiu_15}).
From the boundary points, feature 1.8 is remarked as important by all the models.
It seems that the system sees some relevant information that can be used to discriminate between both groups.
Our clinical experts indicated that feature 1.8 corresponds to the exit zone of the papillomacular bundle.
This bundle is made up of ganglion cell axons which, coming from the most central area of the retina (macula),
reach the temporal border of the optic nerve by means of a practically rectilinear fascicle.
It is therefore highly logical that this is an important feature in determining the presence/absence of the pathology.


\section{Discussion}
\label{sec:Discussion}


The state of the art in ML diagnosis from OCT is made of studies that are focused exclusively
on providing high values of performance while eluding a proper discussion on the usability of the models
and the trust on the decisions made by the proposed solutions.
The major criticism from these works is that these studies have been conducted with a very small number
of samples and the results from the best models are shown without analyzing means, medians, and variances.
These models have low generalization capability and may be establishing unrealistic baselines.
In addition, the methods are used as black-boxes, therefore, we don't know whether the models are taking
decisions in accordance to clinical knowledge that we can trust.
We believe that there is still a long way to go toward solutions for MS diagnosis widely usable and
trustworthy in clinical practice.
To ensure the reliability of the systems, the models should be built up on a large dataset, demonstrate a
high generalization capability, and yield explanations coherent with clinical knowledge.


Our study is a first step toward a trustworthy solution for computer-aided diagnosis of MS from OCT that will
be usable in ophthalmological clinical practice.
Our models are built using the thickness of GCL and RNFL layers, measured with the PPole protocol of Spectralis
SD-OCT devices. Since every subject can contribute to the dataset with two samples, we have analyzed the
performance of the models built with left, right, random, and both eyes.
We did not find any signicative difference in accuracy between using random or both eyes.
We have compared three of the best-performing methods in different applications and challenges with datasets
similar to ours: XGB, RF, and EBM.
Our models confirmed measurements given in PPole protocol being as informative as other measurements in the
discrimination of MS from HC subjects, with performance in the order of the state of the art baseline, with large
sample sizes established in~\cite{Kenney_22}.
We were able to reach an accuracy greater than 0.9 in some folds for some configurations of feature set, sampling,
and methods.
We found that the measurements from the GCL layer were much more useful than RNFL in the discrimination of MS from
HC subjects.
This is in agreement with the latest advances in the state of the art~\cite{LopezDorado_22,Kenney_22}.
From these results, we believe that any of the considered configurations of feature set (Zones or PPole grid),
sampling (rand or LR), and machine learning method could be used as a starting point for a usable computer-aided
solution, once an appropriate large dataset sample has been gathered.


Our study has revealed the combination of XGB or RF and SHAP as a tandem that may perfectly serve the specialist to
make decisions aided by the model explanations.
Global explanations pointed out to zones Z1 and Z2 and points 5.4 and 5.3 (from 64 features), where a visual inspection of the dataset
revealed a break or thinning of the torus shape of the GCL measurements.
We also found that point 3.1 in the boundary of the PPole grid was mostly considered as a relevant feature.
In addition, a reduction of the thickness in these locations was related with a high probability for the MS class.
This is consistent with current knowledge on how MS affects neuroretina damage.

The local analysis of the SHAP values for the failed test samples revealed, for the best-performing models, subjects
with features of the wrongly predicted class that would also confound an expert in most cases.
We also found cases where there is no reason to give a wrong prediction.
In these cases SHAP waterfall plots indicated that the most important features contributed to antagonist decisions,
and the final decision ended up in less relevant feature values.


EBM without interactions obtained an accuracy similar to XGB and RF.
The explanations were given in a similar way to SHAP, which proves that this glass-box method could compete or complement
XGB and RF in the task.
Global explanations also pointed out zones Z1 and Z2 and points 5.4 and 5.3 with the highest scores.
Point 3.1 in the boundary was also considered as a relevant feature together with other boundary points.
It seems that, while feature importance is located over the ridge of the torus shape for XGB and RF, it is spread
out all over the grid for EBM.
Adding pairwise interactions to the model ended up with pairwise explanations as the most relevant features.
The interaction pair 4.5 $\times$ 5.4 was revealed as the most important.
However, the remaining interactions involved boundary features, which may reinforce our suspicions of bias in EBM models.
From these results we believe that EBM may be more sensitive to developing bias than XGB or RF for this kind of data.


%
%
Our clinical experts indicated that a plausible pathophysiological explanation for the importance of feature 3.1 could be based on the
vascular theory that accompanies neurodegenerative diseases, which argues that axonal loss in these pathologies is associated
with a loss of vascular supply, so that both mechanisms (neurodegeneration and vascular loss) are associated in the pathophysiology
of MS.
Feature 3.1 is crossed by medium and large calibre peripapillary blood vessels (see Figure 2).
When the thickness of the GCL is reduced, it is due to the presence of thinner blood vessels and our machine learning methods may be able
to capture this relationship.



From our results, we glimpse a computer-aided solution for the diagnosis of MS where the practitioner is informed with
the PPole grid image, the probability of the ML methods, the local SHAP and EBM explanations, and
some recommendations learned from the waterfall plots in negative cases that would guide the process
of decision making.
In the case of witnessing signs of doubt as the ones shown in this work, complementary examination beyond
the neuroophthalmological eye anatomy or maintaining a wait-and-see approach should be recommended.
This way, OCT can be regarded as a quick test for the assessment of MS, with no side effects, non-invasive and therefore can be
repeated on a serial basis in cases of unclear diagnosis or to see the progression of axonal damage.
OCT is very economical and cost-effective for healthcare systems and can be performed by non-specialised personnel,
making it a tool that in many cases can be used to reduce the number of magnetic resonance imaging requested, and thus alleviate
the healthcare and economic burden on healthcare systems.
Despite computer-aided solutions not able to reach a 100\% accuracy, they would provide a first trustable
assessment in the great majority of cases that would save additional expensive, time-consuming, and invasive tests.



Possible room for improvement lies in the fact that we have worked with PPole grid features as raw data.
The decision was made due to the small sample size and the low resolution of the data.
Given the two-dimensional nature of PPole features, we believe that the performance and explainability may
be improved with models that take into account the spatial representation of the data.
As future work, we will explore convolutional neural networks, vision transformers (ViT), or swim transformers
using the PPole grid features as images.


Although our study shows the methodology to follow to develop this solution, our results need to be confirmed by
a larger sample of patients. Therefore, we are currently working on recruiting more populated and balanced datasets.
We expect that the improved solution will provide models with a good balance between accuracy and explainability.

\section*{Acknowledgments}

This work was partially supported by the national research grants PID2019-104358RB-I00 
(DL-Ageing project), Government of Aragon Group Reference $T64\_20R$ (COS2MOS research group),
Carlos III Health Institute grants PI17/01726 and PI20/00437, and by the Inflammatory Disease Network
(RICORS) (RD21/0002/0050) (Carlos III Health Institute).
Ubaldo Ramon-Julvez work is granted by Government of Aragon.
The funding organizations had no role in the design or conduct of this research.



%
%

\bibliographystyle{spmpsci}      

\bibliography{OCT.bib}

%
%

\end{document}